\documentclass[acmsmall]{acmart}

\AtBeginDocument{%
  }

\usepackage{algorithmic}
\usepackage{graphicx}
\usepackage{textcomp}
\usepackage{xcolor}
\usepackage{adjustbox}
\usepackage{longtable}
\usepackage{booktabs}
\usepackage{colortbl}
\usepackage{tabularx}
\usepackage{multirow}
\usepackage{fontawesome5}
\usepackage{subcaption}
\usepackage{float}
\usepackage{placeins}
\newcommand{\inlineicon}[2][1em]{%
  \adjustbox{valign=B}{\includegraphics[height=#1]{#2}}%
}

\setcopyright{acmlicensed}
\copyrightyear{2025}
\acmYear{2025}
\acmDOI{XXXXXXX.XXXXXXX}
\acmConference[Conference acronym 'XX]{Make sure to enter the correct
  conference title from your rights confirmation email}{June 03--05,
  2025}{Woodstock, NY}
\acmISBN{978-1-4503-XXXX-X/2025/06}

\usepackage[most]{tcolorbox}

\tcbset{
  rqfinding/.style={
    colback=yellow!10,       
    colframe=gray!70!black,
    boxrule=0.8pt,           
    arc=4pt,                 
    left=6pt, right=6pt, top=6pt, bottom=6pt,
    fonttitle=\bfseries,
  }
}

\newtcolorbox{RQFinding}[2][]{
  rqfinding,
  title={#2}, 
  #1
}

\begin{document}

\title{
Collaborative Agents for Automated Program Repair in Ruby
% RAMP: A Lightweight Multi-Agent Program Repair for Ruby with Feedback-Driven LLM
}

\author{Nikta Akbarpour}
\authornotemark[1]
\email{niktakbr@student.ubc.ca}
\author{Mahdieh Sadat Benis}
\authornotemark[1]
\email{mahdiehs@student.ubc.ca}
\author{Fatemeh Hendijani Fard}
\authornotemark[1]
\email{fatemeh.fard@ubc.ca}
\affiliation{%
  \institution{University of British Columbia}
  \country{Canada}
}

\author{Ali Ouni}
\affiliation{%
  \institution{Ecole de Technologie Superieure}
  \country{Canada}}
\email{Ali.Ouni@etsmtl.ca}

\author{Mohamed Aymen Saied}
\affiliation{%
  \institution{Laval university}
  \country{Canada}
}
\email{mohamed-aymen.saied@ift.ulaval.ca}

\begin{abstract}
Automated Program Repair (APR) has advanced rapidly with Large Language Models (LLMs), but most existing methods remain computationally expensive, and focused on a small set of languages. Ruby, despite its widespread use in web development and the persistent challenges faced by its developers, has received little attention in APR research.
In this paper, we introduce RAMP, a novel lightweight framework that formulates program repair as a feedback-driven, iterative process for Ruby. RAMP employs a team of collaborative agents that generate targeted tests, reflect on errors, and refine candidate fixes until a correct solution is found. Unlike prior approaches, RAMP is designed to avoid reliance on large multilingual repair databases or costly fine-tuning, instead operating directly on Ruby through lightweight prompting and test-driven feedback.

Evaluation on the \textsc{XCodeEval} benchmark shows that RAMP achieves a pass@1 of 67\% on Ruby, outperforming prior approaches. RAMP converges quickly within five iterations, and ablation studies confirm that test generation and self-reflection are key drivers of its performance. Further analysis shows that RAMP is particularly effective at repairing wrong answers, compilation errors, and runtime errors.
Our approach provides new insights into multi-agent repair strategies, and establishes a foundation for extending LLM-based debugging tools to under-studied languages.
\end{abstract}


\maketitle

\section{Introduction}

Debugging and fixing software bugs remain some of the most time-consuming and error-prone tasks in software development \cite{weber2025understanding, lee2024unified}. While traditional Automated Program Repair (APR) techniques have shown promise \cite{renzullo2025automated, dikici2025advancements}, recent advances in Large Language Models (LLMs) have opened up new possibilities for more flexible, context-aware solutions \cite{yang2025survey, sun2025empirical, li2025context}. 
However, fine-tuning LLMs is resource intensive \cite{yang2025survey, sun2025empirical, zhang2024systematic, renzullo2025automated, huang2024evolving, zhang2023survey} and 
inference pipelines incur high token costs that make real-world deployment challenging \cite{yang2025survey, zhang2023survey}. 
An alternative avenue is LLM-based agentic flows for APR, which uses multiple agents without fine-tuning to repair a buggy code \cite{luo2025unlocking, ding2024autocoderrover, lee2024unified}. 
Nonetheless, existing works remain narrow. 
Evaluation practices overlook runtime and efficiency, which hinders fair comparison \cite{yang2025survey, zhang2023survey}. 
Moreover, language coverage is highly skewed toward Java, Python, and C, while widely used languages such as Ruby remain largely underexplored ~\cite{zhang2024systematic}.

This gap is notable given Ruby’s wide usage for web applications, adoption in major platforms such as Airbnb, GitHub, and Shopify, and with over two million projects on GitHub~\cite{chandran2020correlative, kaleba2022you}. 
The community of Ruby developers, being more than one million, continue to face persistent challenges related to debugging, test writing, and ensuring application correctness \cite{akbarpour2025unveiling}. Prior studies, drawing on Stack Overflow discussions and developer survey data, reveal recurring pain points~\cite{akbarpour2025unveiling}, emphasizing the need to automated support systems for program repair and debugging. 
Yet, most established APR benchmarks and large-scale evaluations leave Ruby notably absent \cite{zhang2023survey}.

Attempts to broaden APR beyond traditional settings still fall short for Ruby. For example, \textsc{LANTERN} \cite{luo2025unlocking} is notable for being evaluated on \textsc{XCodeEval} benchmark \cite{khan2024xcodeeval} and including Ruby, but its reliance on large multilingual repair databases makes it unsuitable for single-language scenarios with limited data. Meanwhile, \textsc{ChatRepair} \cite{xia2024automated} represents an efficient LLM-based repair approach, but it was not evaluated on Ruby and only provides raw execution results to the model, offering minimal feedback and little support for iterative refinement. These examples highlight the broader challenge: no prior system delivers Ruby-focused, multi-agent, feedback-rich repair.

To address these limitations, we introduce \textbf{RAMP}: \textit{\textbf{R}uby \textbf{A}utomated \textbf{M}ulti-agent \textbf{P}rogram repair}, a lightweight, feedback-driven framework that models APR as an iterative repair loop. RAMP decomposes the APR process into specialized roles, including a \texttt{Programmer Agent}, \texttt{Test Designer}, \texttt{Test Executor}, and \texttt{Feedback Integrator}. Starting from a buggy program and problem context, RAMP leverages specialized agents to reflect on the source of errors, generate guiding tests, propose candidate fixes, and validate them through execution feedback. Candidate repairs are iteratively refined until they either pass the generated tests or the iteration budget is exhausted. 
This multi-agent workflow enables deeper semantic reasoning than single-agent baselines while remaining cost-efficient. Unlike prior systems that rely on large multilingual databases or costly fine-tuning, RAMP operates directly on Ruby code through lightweight prompting and test-driven feedback. By eliminating the need for resource-heavy training pipelines and cross-language translation, it achieves strong repair performance while keeping computation practical for the understudied language, Ruby.

On the \textsc{XCodeEval} benchmark \cite{khan2024xcodeeval}, RAMP attains the best pass@1 on Ruby (67.0\%), outperforming \textsc{LANTERN} (61.7\%) and prompting baselines, and converges by iteration~5. Ablation studies show that test generation and self-reflection are pivotal and removing them reduces the score by 18.1 and 19.3 percentage points, respectively.
In terms of failure categories defined by \textsc{XCodeEval}, RAMP is effective at repairing programs that initially produced WRONG\_ANSWER outputs (i.e., incorrect but executable results, 68.5\% repaired), followed by COMPILATION\_ERROR cases (programs that failed to compile, 66.7\% repaired) and RUNTIME\_ERROR cases (programs that crashed during execution, 60.4\% repaired) but struggles with resource-related failures and advanced categories such as binary search, bitmasks, and matrices. 
These results highlight RAMP’s effectiveness across diverse bug types and its efficiency in converging quickly, underscoring its promise as a practical solution for APR in Ruby.
In summary, our key contributions include:

\begin{itemize}
    \item \textbf{Introducing RAMP}, the first Ruby-focused APR framework that leverages a lightweight, multi-agent design to enable iterative, feedback-driven program repair without relying on large multilingual databases or fine-tuning.
    \item \textbf{Formulating APR as a test-guided repair loop}, where specialized agents collaborate through self-reflection, test generation, candidate repair, and execution feedback, enabling deeper semantic reasoning while remaining cost-efficient.
% \item \textbf{Conducting the first systematic evaluation of APR on Ruby} using the XCodeEval benchmark and identifying the strengths of RAMP. \FHF{Not sure if third item is a contribution.}
    \item \textbf{Releasing all scripts and experimental results} as open source to support replication and future research~\footnote{\url{https://figshare.com/s/829875edc8c876c50de5}}.
    
\end{itemize}

\section{Related Work}
% \FHF{I randomly removed some citations. See if you want to add/replace citations.}

Research on LLM-based APR has progressed along several directions \cite{yang2025survey}: prompting \cite{appatch2024, haque2025towards, xu2024aligning, chen2024large, liu2024t, xia2023automated}, procedural methods \cite{liu2025agent, tang2024code, brancas2025combining, cref2024, xiao2025predicatefix}, fine-tuning approaches \cite{huang2023fine, repairllama2024, ruiz2024lorair, repair2024ppo, wei2025swe, yu2025smart}, and agent-based systems \cite{antoniades2024swe, bouzenia2024repairagent, liu2025agent, su2025learn}.
% Research on LLM-based APR has progressed along several directions \cite{yang2025survey}: prompting \cite{nong2024automated, haque2025towards, xu2024aligning, chen2024large, liu2024t, ouyang2025knowledge, prenner2022can, tian2023chatgpt, xia2022less, gao2023makes, nashid2023retrieval, xia2023automated}, procedural methods \cite{liu2025agent, tang2024code, tao2024magis, yin2024thinkrepair, brancas2025combining, cref2024, hula2023, drcodepilot2024, llm4cve2023, wei2023copiloting, xiao2025predicatefix}, fine-tuning approaches \cite{huang2023fine, jiang2023tenllms, repairllama2024, ruiz2024lorair, huang2024template, xia2023plastic, narrepair2024, bouzenia2023tracefixer, chow2024pyty, jin2023inferfix, repair2024ppo, secrepair2024, swe-rl2024, adapatcher2024, smartllama2024, dai2025less, islam2024llm, wei2025swe, yu2025smart, zhao2024repair}, and agent-based systems \cite{antoniades2024swe, bouzenia2024repairagent, he2025code, li2024cleanvul, zhang2024autocoderover, liu2025agent, luo2025unlocking, su2025learn, tao2024magis, wang2024openhands, yadavally2025large, yang2024swe}.
Prompting-based repair explored how far general-purpose LLMs could be pushed without fine-tuning. AlphaRepair \cite{xia2022less} showed that even zero-shot prompting can produce valid repairs, while evaluations of Codex and ChatGPT \cite{fan2023chatgpt, prenner2022can, tian2023chatgpt} confirmed strong baseline performance but revealed sensitivity to prompt phrasing. Few-shot prompting \cite{xia2023automated, gao2023fewshot, ahmed2023fewshot, nashid2023retrieval} improved stability by providing exemplars, and retrieval-augmented prompting \cite{chen2024large, liu2024t, ouyang2025knowledge} further enriched inputs with repository history, security rules, or knowledge graphs.

% \FHF{What is the difference between procedural and agent systems?} \Nikta{they can have overlap, but not necessarily all agentic systems go through refinement loops and vice versa}
Procedural methods instead adopt iterative loops of generation and testing. \textsc{ChatRepair} \cite{xia2024automated} generates candidate patches, executes them, and re-prompts on failing assertions. ThinkRepair \cite{yin2024thinkrepair} extends this idea by using chain-of-thought reasoning and retaining partially correct patches for reuse. Other systems, such as REx \cite{tang2024code} and ContrastRepair \cite{kong2024contrastrepair}, refine this test–reason–adapt cycle to progressively converge on correct solutions.
Several studies enhance repair with auxiliary signals \cite{wang2023rap, jin2023inferfix, peng2024domain, ntr2024, mansur2024ragfix, yang2025enhancing, xia2024agentless}. TraceFixer \cite{tracefixer2023} leverages execution traces, while graph-based approaches like GLANCE \cite{nashid2023embedding} and SYNSHINE \cite{ahmed2022synshine} capture structural information from control flow or compiler diagnostics. Others exploit developer artifacts such as review comments \cite{zhao2023right}, compiler outputs, or failing test cases \cite{xu2024aligning, appatch2024, haque2025towards}.

% \FHF{You still do not clarify the domain/type of problems RAMP covers, therefore, not clear why other agent systems, like AutoCodeRover and AgentCoder are not considered in your baeslines. }

Autonomous multi-agent frameworks represent the newest wave of APR \cite{bouzenia2024repairagent, liu2025agent, wang2024openhands, wang2024magis}. AutoCodeRover \cite{ding2024autocoderrover} operates on real repositories by navigating ASTs, localizing faults from GitHub issues, and validating patches at the project level. \textsc{LANTERN} \cite{luo2025unlocking} proposes a cross-language paradigm, translating buggy code from weaker to stronger languages based on a decision module that leverages prior repairs from a large multilingual database, which is computationally expensive. Multi-agent pipelines such as MAGIS \cite{wang2024magis}, SWE-Search \cite{antoniades2024swe}, and Learn-by-Interact \cite{su2025learn} emphasize autonomy through planning, search, and debate, while FixAgent \cite{lee2024unified} aligns debugging tasks with cognitive models to unify localization and repair.
Beyond APR-specific work, progress in LLM-based code generation offers useful baselines. Self-Planning \cite{jiang2024self} adopts a two-stage workflow where the model first outlines solution steps via few-shot prompting and then generates code incrementally from the plan. Self-Collaboration \cite{dong2024self} instead distributes roles such as analyst, coder, and tester across multiple LLMs that interact as a virtual development team. Although not designed for repair, both approaches emphasize planning and collaboration, which are also critical in APR, though neither has been evaluated on Ruby.

Despite these advances, language coverage remains skewed toward Java, Python, and C, while Ruby accounts for only about 1\% of studies~\cite{zhang2024systematic}.
The existing approaches for APR in Ruby depend on multilingual datasets~\cite{luo2025unlocking}, overlooking runtime and efficiency metrics~\cite{renzullo2025automated, huang2024evolving}. 
The reliance on code translation~\cite{luo2025unlocking} makes it unsuitable for single-language settings with limited data. 
Other works~\cite{xia2024automated}, while efficient, only return raw execution results to the model rather than generating richer feedback to guide repair.
% ChatRepair \cite{xia2024automated}, while efficient, only returns raw execution results to the model rather than generating richer feedback to guide repair. Self-Planning \cite{jiang2024self} and Self-Collaboration \cite{dong2024self}, although conceptually related, were developed for code generation rather than repair and have not been tested on Ruby.}
In this work, we address these gaps by introducing \textsc{RAMP}, a reflection-augmented multi-agent pipeline tailored to Ruby. Unlike previous studies~\cite{luo2025unlocking, xia2024automated}, \textsc{RAMP} avoids the need for multilingual databases and integrates explicit feedback generation, rather than simply passing back test results.
While systems such as AutoCodeRover \cite{ding2024autocoderrover} target repository-scale repair, due to benchmark existence, our focus is on competitive-programming style tasks: short, single-file programs with well-defined I/O-based test cases, as captured in the \textsc{XCodeEval} benchmark. Within this domain, relevant baselines are agentic methods that, like \textsc{RAMP}, operate on self-contained snippets and optimize for functional correctness rather than project-level integration.
Our design emphasizes a balanced trade-off between resource efficiency and accuracy.

\section{Methodology}

\subsection{Methodology Overview}
\label{problem_setup}
We formulate the problem of APR using RAMP as follows.
For each buggy instance we are given: problem context $C$ (natural-language specification), sample input and output pairs $S$, buggy program $d$ (source code to be repaired), hidden benchmark tests $T_h$, and iteration budget $K$.
A \emph{candidate repaired program} is denoted by $r$.
Generated tests in RAMP are denoted as $T_g$ and are used for guidance during repair generation.

\paragraph{Objective.}
Given a problem $C$ and input-output pairs $S$ for a buggy program $d$, the objective is to return a repaired program $r$ that passes all hidden tests $T_h$ within the iteration budget $K$.
Formally:
\[
\textsc{RAMP}(C,S,d,T_h,K) =
\begin{cases}
r_t & \text{if } \exists\, t \le K \text{ such that } 
\Big( \mathrm{Eval}(r_t,T_g) = \texttt{pass} 
      \;\;\lor\;\; t=K \Big) 
      \;\wedge\; \mathrm{Eval}(r_t,T_h) = \texttt{pass}, \\[6pt]
\textsc{Fail} & \text{otherwise.}
\end{cases}
\]
Here, $r_t$ is the $t$-th candidate program, and hidden tests $T_h$ are only executed if either (i) $r_t$ passes all generated tests $T_g$, or (ii) the iteration budget $K$ is exhausted.

\begin{figure*}[h]
    \centering
    \includegraphics[width=\textwidth]{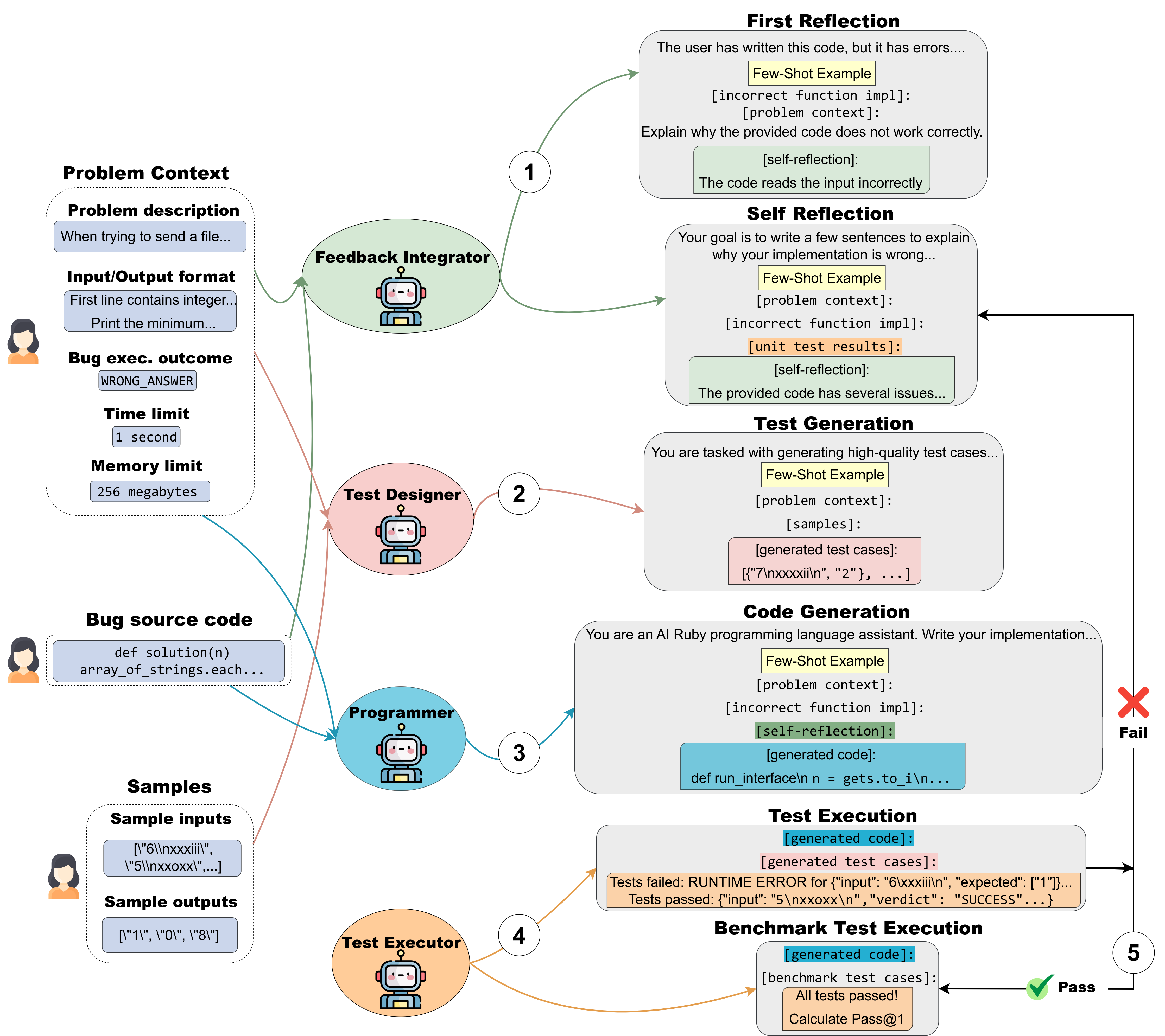}
    \caption{Overview of the RAMP framework.
    \inlineicon{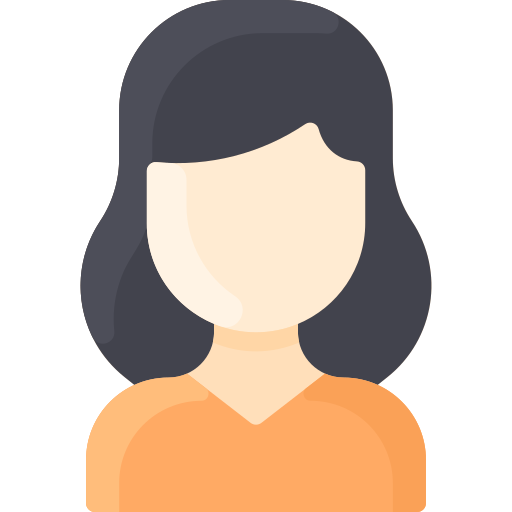} denotes benchmark-provided inputs (problem description, I/O format, sample I/O, limits, and buggy code); 
    \inlineicon{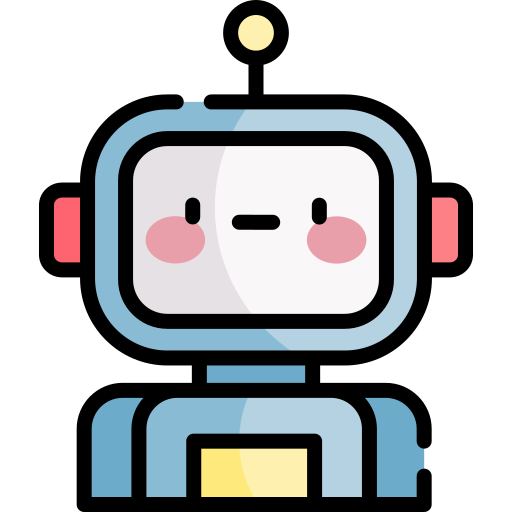} denotes agents. 
Numbered stages: (1) \emph{Feedback Integrator} produces a natural language self-reflection; 
(2) \emph{Test Designer} generates the public test cases $T_g$; 
(3) \emph{Programmer} generates full repaired code candidates; 
(4) \emph{Test Executor} runs candidates on $T_g$ and returns results; 
(5) candidates that either pass $T_g$ or exhaust the iteration budget are validated on hidden tests $T_h$, and success requires passing all $T_h$ (we report pass@1). 
If a candidate \emph{fails} $T_g$, the process continues iteratively, repeating (1), (3), and (4) with updated reflection until $T_g$ passes or the iteration budget $K$ is reached. 
Arrow colors show data flow and match the corresponding agent colors; highlighted text blocks use the same color to indicate which agent produced that output.}
    \label{fig:methodology_overview}
\end{figure*}

Figure~\ref{fig:methodology_overview} illustrates how the abstract formulation above is realized in practice through an iterative workflow with five main steps coordinated by four specialized agents:

\textbf{Step~1: Initial Reflection (Feedback Integrator Agent).}  
    The process begins with the \textit{Feedback Integrator Agent}, which produces a natural-language hypothesis $e_0 = R_f(C,d)$ about the potential cause of the bug. At this stage, the LLM is prompted to generate only a natural language explanation of why the code may be incorrect, without producing any code. This reflection highlights discrepancies between the intended specification and the observed behavior, and serves as structured guidance for subsequent repair attempts (see Section~\ref{feedback_integrator} for details).

\textbf{Step~2: Test Case Generation (Test Designer Agent).}  
    The \textit{Test Designer Agent} generates a set of guiding test cases $T_g = G(C,S)$, typically including two examples each of basic, edge, and large-scale inputs (total six test cases per input). These tests serve as the basis for execution feedback and are the first barrier that candidate repairs must pass (see Section~\ref{test_designer} for details).

\textbf{Step~3: Candidate Repair (Programmer Agent).}
At each iteration $t$, the \textit{Programmer Agent} produces a candidate repair program $r_t = P(C, d, e_t)$, where $C$ is the problem context, $d$ is the buggy code, and $e_t$ is the reflection generated in Step~1. The agent is prompted not only to generate code but also to reason explicitly about the bug before proposing a fix (see Section~\ref{programmer} for details). 

\textbf{Step~4: Execution and Feedback (Test Executor Agent + Feedback Integrator Agent).}  
    The \textit{Test Executor Agent} runs the candidate $r_t$ on the generated tests $T_g$ from Step~2, producing verdicts and traces $(\text{verdict}_t, \tau_t)$ (see Section~\ref{test_executor} for details).  
    \begin{itemize}
        \item If $r_t$ fails some $T_g$, the \textit{Feedback Integrator Agent} uses the error traces $\tau_t$ to update the reflection $e_{t+1} = R_f(C,d,\tau_t)$, guiding the next repair attempt. The loop then continues with a new candidate (Step~1).  
        \item If $r_t$ passes all $T_g$, the system triggers hidden validation (Step~5).  
    \end{itemize}
    The loop continues until either a candidate passes $T_g$ or the budget $K$ is reached.

\textbf{Step~5: Benchmark Validation (Test Executor Agent).}  
    Hidden benchmark tests $T_h$ are executed only under two conditions: (i) a candidate repair code $r_t$ passes all generated tests $T_g$, or (ii) the maximum iteration budget $K$ is exhausted. The system then returns the first $r_t$ such that $\mathrm{Eval}(r_t,T_h) = \texttt{pass}$. If no candidate satisfies this, the output is \textsc{Fail}.
Though using the hidden tests' execution outcomes boosts the performance of RAMP significantly (see Section~\ref{sec:discussion}), we use benchmark test cases only for the final validation. This ensures the practicality of RAMP in real world, where test cases are not always available. 

In the following, we explain each agent in detail. Due to space limitations, the prompts used for each LLM-based agent are included in the replication package. 

\subsubsection{Feedback Integrator Agent}
\label{feedback_integrator}
To enhance repair accuracy and interpretability, we experiment with two strategies for generating self-reflection from the problem context and buggy source code.  

The first strategy, \textit{specification understanding}, incorporates specification-driven reasoning. It decomposes reflection into three stages: (1) the model infers a functional specification from the natural language problem description, (2) it explains the behavior of the buggy code through step-by-step reasoning, and (3) it compares the intended and observed behaviors to identify the source of the bug.  
The second strategy, which we call \textit{direct error reasoning}, skips explicit specification inference. Instead, the model directly analyzes the buggy code in the context of the problem description to explain why the implementation is incorrect and to detect the likely source of the error.  

By comparing these two strategies, we investigate whether specification-driven reasoning enables more effective bug localization and repair guidance than relying on direct reasoning over the buggy code and problem context alone. The results of this comparison are presented in the ablation study for RQ2 (see section \ref{RQ2}). 
Building on this foundation, the \textit{Feedback Integrator Agent} plays a central role in the repair loop. It is responsible for (1) producing the initial self-reflection using one of the above strategies, and (2) integrating signals from the \textit{Test Executor} to guide iterative repair. When a candidate solution fails, the agent examines execution traces, error logs, and discrepancies between expected and actual outputs, and produces a concise natural language summary. This structured feedback is then provided to the \textit{Programmer Agent}, informing the next repair attempt.

\subsubsection{Test Designer Agent}
\label{test_designer}
The \textit{Test Designer Agent} is responsible for generating the test cases used to evaluate the functional correctness of candidate repairs. These cases not only verify correctness but also form the basis for producing meaningful feedback during the iterative repair process.  

Rather than relying on the hidden unit tests provided in the benchmark, we generate our own test cases for two main reasons. First, the benchmark includes a large number of hidden tests, and executing them after every repair attempt would be computationally expensive. By contrast, our approach evaluates candidate solutions on a fixed set of only six test cases, which greatly reduces execution cost. Second, while the hidden test suites do cover diverse scenarios, selectively sampling from them to ensure balanced representation across categories is non-trivial. By generating our own test cases, we directly enforce such diversity through a controlled design.  

To ensure broad coverage, we prompt the agent to produce test cases across three complementary categories: (i) \textit{basic cases} to verify general correctness, (ii) \textit{edge cases} to capture boundary conditions and unusual behaviors, and (iii) \textit{large-scale cases} to assess performance and scalability. For each category, the agent generates two test cases, resulting in a compact yet diverse suite of six, designed to remain lightweight.  
For this purpose, the agent produces input–output pairs in a single step. These generated cases are then passed to the \textit{Test Executor} for evaluation against candidate repairs.

\subsubsection{Programmer Agent}
\label{programmer}
The \textit{Programmer Agent} is implemented using an LLM and is responsible for generating candidate repairs. It receives the problem context, buggy code, and prior reflections, and produces corrected implementations of the buggy code.  
We experiment with two prompting strategies: (1) \textit{Chain-of-Thought (CoT) few-shot prompting}~\cite{wei2022chain}, which provides intermediate reasoning steps, and (2) \textit{Structured Chain-of-Thought (SCoT) prompting}~\cite{li2025structured}, which decomposes the task into subtasks such as analyzing the bug, outlining a repair plan, and then producing the corrected code. SCoT explicitly organizes intermediate reasoning around core program structures (sequence, branch, and loop), encouraging the model to think in terms of how source code is constructed and thereby improving the reliability and quality of generated repairs~\cite{li2025structured}.
The results of this comparison are presented in the ablation study for RQ3 (see section \ref{RQ3}).

In addition, we investigate the effect of each prompt components on repair performance. Specifically, we examine how results change when (i) including sample inputs and outputs from the benchmark in the prompt, (ii) removing resource-related constraints such as time and memory limits, and (iii) omitting descriptions of input and output structure. These variations allow us to assess which contextual signals strongly influence the model’s ability to generate correct repairs.  

Since initial repairs may still contain errors (syntax issues, logical flaws, or failed tests), the \textit{Programmer Agent} iteratively refines its solutions using structured feedback from the \textit{Feedback Integrator} until a valid repair is achieved or the iteration limit is reached.

\subsubsection{Test Executor Agent}
\label{test_executor}
The \textit{Test Executor Agent} is a non-LLM component implemented as a Python script that executes Ruby code in a controlled runtime environment. It provides reliable execution-based validation of candidate repairs.  
Given candidate code from the \textit{Programmer} and test cases from the \textit{Test Designer}, the executor runs the program against each case and captures outputs, exceptions, and exit status. If all cases pass, the repair is marked correct and advanced to benchmark validation. If any test fails, the executor collects error messages and runtime traces, which are then passed to the \textit{Feedback Integrator Agent} to produce feedback. This enables the iterative cycle of reflection, repair, and re-execution that underpins the RAMP framework.

\subsection{Benchmark}

To evaluate our APR system for Ruby, we adopt the \textsc{xCodeEval} benchmark~\cite{khan2024xcodeeval}, a large-scale, multilingual, and multitask dataset for evaluating code understanding, generation, translation, and retrieval capabilities of LLMs. \textsc{xCodeEval} provides execution-based evaluation across $26$ programming languages and multiple tasks, making it a comprehensive testbed for cross-lingual and functional code evaluation.
For our task, following the methodology used by recent studies~\cite{luo2025unlocking}, we focus on the validation set of the \textit{APR} task in \textsc{xCodeEval}, which includes real-world buggy code snippets and their corresponding fixed versions. This validation set contains 5,068 samples in total across 11 languages, with 343 Ruby samples.

Each instance in this subset includes a buggy function implementation, a natural language problem description, a set of input–output unit tests, the reference correct solution, the bug’s execution outcome, and additional meta-information such as problem tags that specify the type of solution approach required (e.g., graphs, sorting), and the difficulty level.

\subsection{Model and Experimental Setup}

We utilize two state-of-the-art instruction-tuned code language models: DeepSeek-Coder 6.7B-Instruct \cite{deepseek2025} and Qwen2.5-Coder-7B-Instruct \cite{hui2024qwen2}. To enable efficient execution under limited GPU resources, we apply 4-bit quantization, which significantly reduces memory consumption. These models are integrated across all the agents except \textit{Test Executor}. For a fair comparison, we also adopt DeepSeek-Coder 6.7B-Instruct as the backbone model for all baseline methods. Our selection is informed by recent technical reports demonstrating the strong performance of these models on code generation and reasoning benchmarks \cite{deepseek2025,hui2024qwen2}, establishing them among the leading open-source models for software engineering tasks. 

For the RAMP experiments with DeepSeekCoder, we used an NVIDIA Tesla V100 GPU equipped with 32GB memory. For all Qwen experiments as well as the baseline methods, we used an NVIDIA H100 SXM5 GPU with 80GB memory. 
The choice of hardware was driven by the resource requirements of each setup. RAMP with DeepSeekCoder is relatively lightweight and can be executed efficiently on a V100, while Qwen and baseline methods require larger memory and compute capacity, making the H100 a more suitable option. We ran RAMP and all iterative baselines for 11 iterations, following the setup of a recent baseline, LANTERN~\cite{luo2025unlocking}, which requires translation to 11 different programming languages. In \textsc{RAMP}, we set the temperature to $0.8$ for code generation and $0.1$ for all other agents. We apply sampling with $top\_p = 0.95$ and generate a single candidate per bug. 
For baseline methods, we use the hyperparameters specified in their original implementations.

\subsection{Evaluation Metric}
To assess repair quality, we adopt the \textit{pass@k} metric, which quantifies the chance that a model generates at least one valid solution within its top-$k$ outputs \cite{kulal2019spoc}. Given $n$ candidate fixes for a problem, with $c$ of them correct, pass@k is defined as:

\begin{equation}
\text{Pass@}k = \mathbb{E}_{\text{problems}} \left[ 1 - \frac{\binom{n-c}{k}}{\binom{n}{k}} \right],
\end{equation} 

We adopt \textbf{pass@1} \cite{kulal2019spoc}, to balance evaluation fidelity with computational efficiency, as focusing on the top-ranked candidate reduces resource consumption while still providing a meaningful indicator of model performance.
We report pass@1 with a greedy approach, where only one sample is generated, which is equivalent to the percentage of solved problems. As this is a deterministic generation with no randomness, the results are comparable without requiring statistical tests.

\subsection{Baselines}
We compare our approach against a diverse set of recent LLM-based APR baselines. In particular, we evaluate six representative methods: LANTERN~\cite{luo2025unlocking}, \textsc{ChatRepair}~\cite{xia2024automated}, Self-Planning~\cite{jiang2024self}, Self-Collaboration~\cite{dong2024self}, as well as Few-Shot and Zero-Shot prompting. 

\textbf{LANTERN}~\cite{luo2025unlocking} is a state-of-the-art framework that repairs programs by translating the code into another language, fixing it, and then back-translating it, while incorporating iterative feedback to improve patch correctness. We include LANTERN in our comparison because it is one of the most recent approaches, relies on the same benchmark (\textsc{xCodeEval}), and, to the best of our knowledge, is the only framework that reports results for Ruby APR. Since its method depends heavily on cross-language translation, we follow their setup by including all 11 languages they used, with up to 11 iterations. Due to resource constraints, we evaluate on a 10\% subset of the \textsc{xCodeEval} validation set, sampled to preserve the original language and difficulty distribution of the benchmark. For consistency, we run all baselines and RAMP for 11 iterations on the same 10\% subset.

\textbf{ChatRepair}~\cite{xia2024automated} refines candidate repairs over multiple iterations by engaging in conversational feedback with an LLM. A key point to note is that \textsc{ChatRepair} assumes perfect fault localization, meaning that the location of the bug is provided to the system through external tools.

\textbf{Self-Planning}~\cite{jiang2024self} introduces an explicit reasoning step, enabling the model to first outline a repair strategy before generating code.

\textbf{Self-Collaboration}~\cite{dong2024self}, on the other hand, simulates a group of interacting agents who work together to refine candidate patches.

Although Self-Planning and Self-Collaboration were originally developed for code generation rather than APR, we include them here for two reasons: (i) both address reasoning and collaboration, which are central to repair, and (ii) they were chosen as baselines in the LANTERN~\cite{luo2025unlocking} replication package, making their inclusion important for comparability with prior work.
Finally, we consider two widely used prompting settings: \textbf{Zero-Shot}, where the model generates repairs without examples, and \textbf{Few-Shot}, where we provide a single example to guide the repair process.

\section{Results}

To evaluate RAMP's performance and analyze it across different dimensions, our experiments focus on the following research questions: 
% To evaluate RAMP’s effectiveness, analyze the contribution of its components, and assess its performance across different dimensions of the repair task, our experiments focus on the following research questions:

\textbf{RQ1: How does RAMP perform in comparison to other APR methods?}

\textbf{RQ2: How do the different agents in RAMP contribute to its overall effectiveness?}  

\textbf{RQ3: How do prompting strategies and refinement iterations influence RAMP’s repair performance?}  

\textbf{RQ4: How does RAMP perform across tasks of varying difficulty, subject domains, and initial execution outcomes?}

Due to resource limitations, we used a 10\% sampled subset of the dataset for all experiments in RQ1 (including RAMP and all the baselines) corresponding to 34 questions for Ruby. For all other research questions, which are evaluated only on RAMP, we use the full validation set.
Additionally, for all experiments including RAMP and baselines (except the left plot of Figure~\ref{fig:side_by_side}), we report cumulative pass@1, where the total number of tasks that have been solved at least once up to a given iteration is summed across all prior iterations and reported. 

\subsection{RQ1: RAMP Performance Compared to Other APR Methods}

To assess the effectiveness of \textsc{RAMP}, we compare its performance with the pass@1 score on the Ruby subset of the benchmark against existing APR baselines as reported in Table~\ref{tab:apr_compare}. Among all methods, \textsc{RAMP} achieves the highest pass@1 score of 67.0\%, substantially outperforming prior approaches. For instance, \textsc{LANTERN}, the strongest baseline, attains 61.7\%, while Self-Planning reaches 56.0\%, Few-Shot achieves 47.5\%, and Zero-Shot drops to only 24.1\%. Other methods, such as \textsc{ChatRepair} (17.6\%) and Self-Collaboration (0.0\%) perform considerably worse. The results demonstrate that \textsc{RAMP} delivers a clear performance improvement over both instruction-tuned prompting baselines and specialized APR systems.

\newcolumntype{C}{>{\centering\arraybackslash}X}
\begin{table}
\centering
\begin{tabularx}{\textwidth}{l*{7}{C}}
\toprule
\textbf{Language} & \textbf{Zero-Shot} & \textbf{Few-Shot} & \textbf{Self-Planning} & \textbf{LANTERN} & \textbf{Chat-Repair} & \textbf{Self-Collab.} & \textbf{RAMP} \\
\midrule
% C           & -- & -- & 47\% & 6.6\% & 10.5\% & 20\% & --  \\
% C\#         & -- & -- & 24\% & 26.6\% & 6\% & 11\% & -- \\
% C++         & -- & -- & 26\% & 22.7\% & 9\%  & 7\% & -- \\
% Go          & -- & -- & 19\% & 52.8\% & 18\% & 14\% & -- \\
% Java        & -- & -- & 21\% & 39\% & 8\% & 11\% & -- \\
% JavaScript  & -- & -- & 61\% & 70\% & 27.7\% & 54\% & -- \\
% Kotlin      & -- & -- & 23\% & 48\% & 22.3\% & 8\% & -- \\
% PHP         & -- & -- & 24\% & 50\% & 14.4\% & 7\% & -- \\
% Python      & -- & -- & 27\% & 31.2\% & 8\% & 5\% & -- \\
Ruby        & 24.1\% & 47.5\% & 56.0\% & 61.7\% & 17.6\% & 0.0\% & \textbf{67.0\%} \\
% Rust        & -- & -- & 39\% & 28.5\% & 20.8\% & 5\% & -- \\
\bottomrule
\end{tabularx}
\caption{Pass@1 for different APR approaches on Ruby language.}
\label{tab:apr_compare}
\end{table}

\begin{figure}
    \centering
\begin{subfigure}[t]{0.48\columnwidth}
        \centering
        \includegraphics[width=\linewidth]{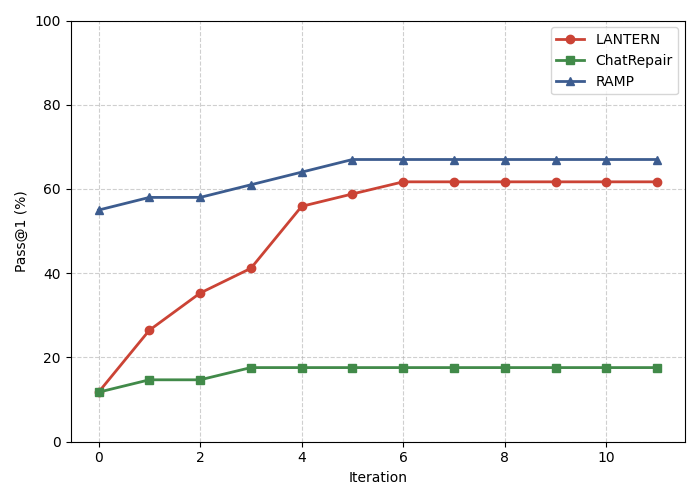}
        \label{fig:bug_outcome}
    \end{subfigure}
    \hfill
        \begin{subfigure}[t]{0.48\columnwidth}
        \centering
        \includegraphics[width=\linewidth]{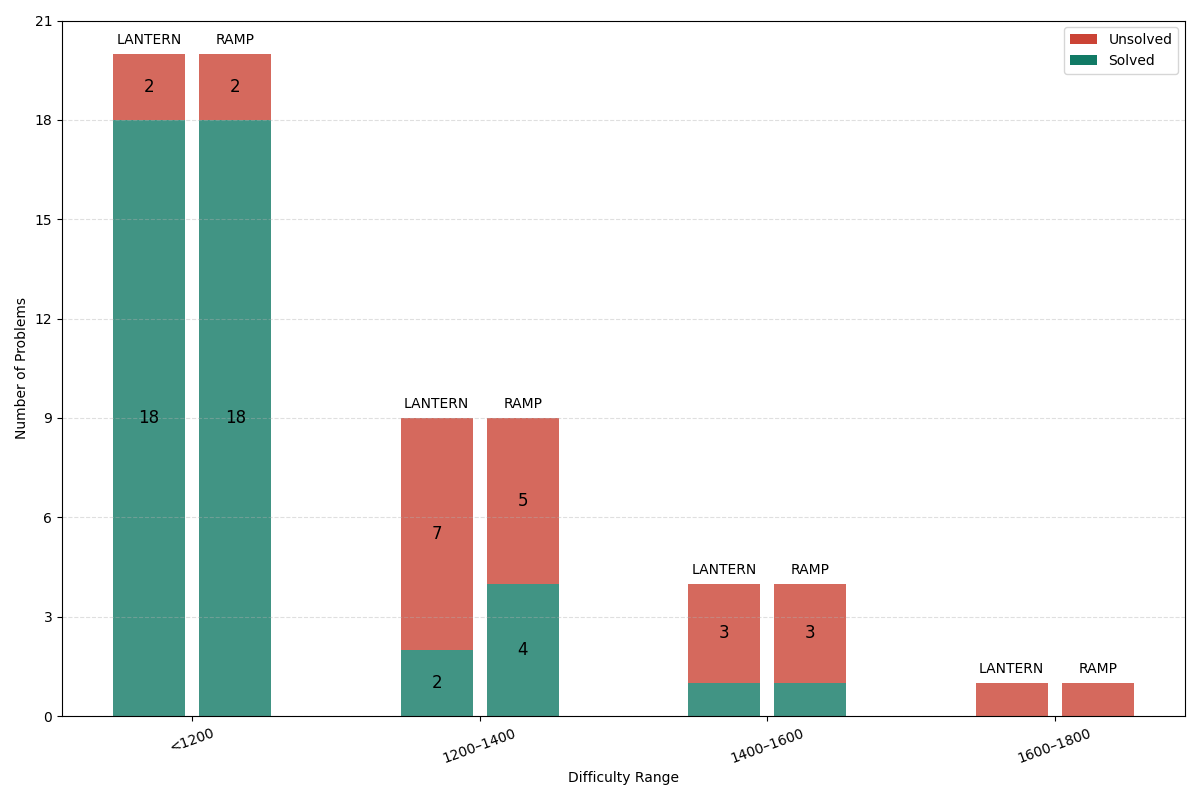}
        \label{fig:difficulty}
    \end{subfigure}
    
    \caption{\textbf{Left:} Pass@1 of RAMP, \textsc{LANTERN}, and \textsc{ChatRepair} over iterations.
    \textbf{Right:} Distribution of solved and unsolved problems after applying RAMP and \textsc{LANTERN} in different difficulty ranges.}
    \label{fig:ramp_vs_lantern_diff_base}
\end{figure}

We further analyze the evolution of repair performance across iterations in the left plot of Figure~\ref {fig:ramp_vs_lantern_diff_base}. \textsc{RAMP} exhibits rapid improvements in the early stages: performance rises from 55.0\% at iteration 0 to 67.0\% by iteration five, after which results plateau. In contrast, \textsc{LANTERN} starts from a much lower baseline of 11.7\% and requires seven iterations to converge at 61.7\% Similarly, \textsc{ChatRepair} shows negligible gains over time, remaining near 17.6\% throughout. 
This highlights that \textsc{RAMP} not only achieves the highest eventual accuracy but also \textit{converges significantly faster} than existing approaches. 
Additionally, in contrast to other baselines, \textsc{RAMP} design enables relatively high repair performance at the early iterations, enabling developers to obtain correct solutions for a higher number of problems, specifically when the iteration budget is low.  
% enabling developers to obtain correct solutions with fewer attempts. This early effectiveness reduces computational cost and turnaround time, making \textsc{RAMP} more efficient and practical for real-world debugging scenarios.

The right plot of Figure \ref{fig:ramp_vs_lantern_diff_base} compares the distribution of solved and unsolved problems across difficulty ranges for RAMP and the baseline with highest score, \textsc{LANTERN}.
Both systems perform best on easier tasks ($<1200$), solving 18 cases each and leaving only two unsolved. In the 1200--1400 range, RAMP achieves stronger performance with four solved and five unsolved, compared to \textsc{LANTERN}’s two solved and seven unsolved. At higher difficulty levels, the success rate drops for both systems.
% : in the 1400--1600 range, each solves only a small fraction of problems, and beyond 1600 almost no tasks are solved. 
Overall, the figure highlights that RAMP maintains a slight advantage over \textsc{LANTERN} at medium difficulty levels, while both approaches show similar strengths on easy tasks and face clear limitations on the hardest problems.

\begin{RQFinding}{Key Finding for RQ1}
\textsc{RAMP} achieves the highest pass@1 on Ruby (67.0\%), substantially outperforming all evaluated baselines such as \textsc{LANTERN} (61.7\%), \textsc{Self-Planning} (56.0\%), and \textsc{ChatRepair} (17.6\%). Across iterations, \textsc{RAMP} reaches strong performance quickly and stabilizes, unlike baselines.
% whereas baselines plateau at much lower levels. 
Moreover, \textsc{RAMP} solves more problems than \textsc{LANTERN}, with a clear advantage on medium tasks, though both methods struggle on the hardest problems.
\end{RQFinding}

\subsection{RQ2: Contribution of Different Agents in RAMP}
\label{RQ2}

In this RQ, we assess the contribution of test generation, self-reflection, first reflection, reflection (self-reflection and first reflection), and inferring specifications to the RAMP framework.
Table \ref{tab:ramp_variants} reports the effect of removing different components of the framework, or adding inferring specifications to RAMP. We conduct these ablation experiments on the two studied LLMs.
Removing test generation forces the framework to create self-reflections solely from the problem context and generated code, without feedback from execution results. This substantially degrades performance, with pass@1 dropping from 66.5\% to 48.4\% ($-18.1$ points) for DeepSeekCoder, while Qwen Coder remains largely unchanged (55.7\% to 56.2\%). The larger decline for DeepSeekCoder suggests that it depends more heavily on execution-based feedback to guide repairs.

\begin{table}[htbp]
\centering
\setlength{\tabcolsep}{3pt} % tighten column padding
\begin{tabularx}{\textwidth}{l*{6}{>{\centering\arraybackslash}X}}
\toprule
\textbf{Pass@1} & \multicolumn{6}{c}{\textbf{RAMP variants}} \\
\cmidrule(l){2-7}
& \shortstack{\scriptsize w/o Test Gen.}
& \shortstack{\scriptsize w/o Self-Refl.}
& \shortstack{\scriptsize w/o First Refl.}
& \shortstack{\scriptsize w/o Reflection}
& \shortstack{\scriptsize w Spec.}
& \shortstack{RAMP} \\
\midrule
\textbf{DeepSeekCoder} & 48.4\% & 47.2\% & 50.4\% & 50.7\% & 63.8\% & 66.5\% \\
\textbf{Qwen Coder}    & 56.2\% & 56.2\% & 54.5\% & 54.5\% & 57.1\% & 55.7\% \\
\bottomrule
\end{tabularx}
\caption{Effect of components in RAMP with pass@1 accuracy. w/o means without, and w means with. }
\label{tab:ramp_variants}
\end{table}

\begingroup
\begin{table}[!ht]
\centering
% \begin{table}
% \centering
\begin{tabular}{c|rrrr}
\toprule
\textbf{Iteration} & \textbf{FN (\%)} & \textbf{FP (\%)} & \textbf{TN (\%)} & \textbf{TP (\%)} \\
\midrule
1  & 185 (53.9\%) & 9 (2.6\%)   & 146 (42.6\%) & 3 (0.9\%) \\
2  & 187 (54.5\%) & 10 (2.9\%)  & 146 (42.6\%) & 0 (0.0\%) \\
3  & 192 (56.0\%) & 0 (0.0\%)   & 151 (44.0\%) & 0 (0.0\%) \\
4  & 196 (57.1\%) & 1 (0.3\%)   & 145 (42.3\%) & 1 (0.3\%) \\
5  & 195 (56.9\%) & 0 (0.0\%)   & 148 (43.1\%) & 0 (0.0\%) \\
6  & 201 (58.6\%) & 0 (0.0\%)   & 142 (41.4\%) & 0 (0.0\%) \\
7  & 203 (59.2\%) & 0 (0.0\%)   & 140 (40.8\%) & 0 (0.0\%) \\
8  & 195 (56.9\%) & 0 (0.0\%)   & 148 (43.1\%) & 0 (0.0\%) \\
9  & 204 (59.5\%) & 2 (0.6\%)   & 137 (39.9\%) & 0 (0.0\%) \\
10  & 204 (59.5\%) & 0 (0.0\%)   & 139 (40.5\%) & 0 (0.0\%) \\
11 & 204 (59.5\%) & 0 (0.0\%)   & 139 (40.5\%) & 0 (0.0\%) \\
\bottomrule
\end{tabular}
\caption{Distribution of generated test case outcomes across iterations: Number of tests falling into each category (Proportion of tests within each iteration that belong to that category).}
\label{tab:test_confusion}
\end{table}
\endgroup

When self-reflection is removed, pass@1 decreases by 19.3 points for DeepSeekCoder, but Qwen Coder remains largely unchanged (55.7\% to 56.2\%), highlighting that structured reflection is substantially more important for DeepSeekCoder than for Qwen Coder. A similar trend holds when the first reflection is removed, where performance falls by 16.1 points for DeepSeekCoder and 1.2 points for Qwen Coder. Interestingly, removing both self-reflection and first reflection leads to a 15.8 points decline for DeepSeekCoder and just 1.2 points for Qwen Coder, further emphasizing the limited reliance of Qwen Coder on reflective mechanisms compared to DeepSeekCoder.
Based on the results in Table~\ref{tab:ramp_variants}, we use the DeepSeekCoder model for all subsequent experiments.

Finally, we test an alternative reflection strategy based on inferred specifications. Instead of relying solely on few-shot CoT prompting, the model is explicitly asked to (1) restate the problem description in its own words, (2) describe the functionality of the buggy code, and (3) identify discrepancies between the two. This structured inference reduces performance for DeepSeekCoder (63.8\%, -2.7 points) but slightly improves Qwen Coder (57.1\%, +1.4 points). These results indicate that while inference-driven reflections capture semantic mismatches, they may over-complicate the reasoning process and ultimately hinder repair effectiveness.

To further evaluate the reliability of generated test cases, we examine whether their expected outputs in generated test cases are consistent with the given inputs and the problem context. 
For this purpose, we adopt a similar experiment as done in previous works \cite{shinn2023reflexion}.
As the \textsc{xCodeEval} benchmark provides corrected ground-truth solutions for each sample, we use these as an oracle to verify the correctness of generated tests. Specifically, we execute the generated test cases against the ground-truth implementations and categorize the outcomes as follows:

\begin{itemize}
    \item \textbf{True Positive (TP):} The generated test cases pass on the ground-truth code \emph{and} the generated code also passes on hidden unit tests from the benchmark.
    \item \textbf{False Negative (FN):} The generated test cases fail on the ground-truth code, \textit{but} the generated code nevertheless passes on the benchmark's hidden unit tests.
    \item \textbf{False Positive (FP):} The generated test cases pass on the ground-truth code, \textit{but} the generated code fails on hidden unit tests from the benchmark.
    \item \textbf{True Negative (TN):} The generated test cases fail on the ground-truth code \textit{and} the generated code also fails on the benchmark's hidden unit tests.
\end{itemize}

Among these categories, \textbf{FNs are less harmful than FPs}. In FN cases, the framework may still recover by producing a correct repair, as the generated code eventually succeeds on hidden tests despite the misleading test outcome. By contrast, in FP cases, the framework incorrectly assumes success based on the generated tests, proceeds to final evaluation, and inevitably fails on hidden test cases. This makes FPs particularly detrimental, as they prevent the framework from exploring further repair opportunities.

The results are shown in Table \ref{tab:test_confusion}. The majority of cases fall into FN and TN categories, with FNs steadily increasing over iterations (from 53.9\% to 59.5\%), while FPs remain consistently rare ($<3\%$). The near absence of TPs indicates that generated test cases rarely align perfectly with both the ground truth and hidden test cases.

These results suggest that while most generated tests are conservative (either failing with ground truth or leading to recoverable false negatives), the presence of FP cases is very low, showing the effectiveness of the generated test cases in RAMP.

\begin{RQFinding}{Key Finding for RQ2}
\textsc{RAMP}’s performance relies heavily on test generation and reflection, especially for DeepSeekCoder (up to 19.3 points drop without them), while Qwen Coder is far less sensitive ($<1.5$ points).  
Specification-inference reflections give mixed results, slightly helping Qwen (+1.4 points) but reducing DeepSeekCoder (-2.7 points).  
Generated tests are generally reliable: FPs are rare ($<3\%$) and FNs, though common, are less damaging since correct repairs can still emerge.
\end{RQFinding}

\subsection{RQ3: Effect of Prompting Strategies and Refinement Iterations on RAMP’s Performance}
\label{RQ3}

In this research question, we investigate how different prompting strategies affect code generation performance. We begin with \textit{Structured Chain-of-Thought (SCoT)} prompting, which guides the LLM to produce structured intermediate reasoning steps during code generation~\cite{li2025structured}. Prior work has reported that SCoT outperforms standard CoT prompting for code-related tasks~\cite{li2025structured}. Using few-shot SCoT prompting, we obtain the pass@1 results shown in Table~\ref{tab:prompting_overview}. Next, we replace SCoT with few-shot CoT prompting (noted as RAMP in Table~\ref{tab:prompting_overview}). Interestingly, this change yields a 2.1-point improvement in pass@1. Based on this empirical observation, we adopted few-shot CoT prompting as our primary strategy for code generation, as reported in all other results.

Next, we investigate which components of the benchmark data should be included in the prompt to improve the performance of the Programmer Agent. The results of this ablation study are shown in Table~\ref{tab:prompting_overview}. First, when input and output specifications are removed from the prompt, the pass@1 score decreases by 10.5 points compared to the setting where they are included. Removing the time and memory limits also reduces performance, with pass@1 dropping from 66.5\% to 64.7\%. Finally, when we add sample input--output pairs to the prompt, the pass@1 score decreases by 2.7 points; possibly because the additional examples increase prompt length and complexity, leading the model to overfit to specific cases or become distracted from the general problem description.

\begin{table}[!htbp]
\centering
\begin{tabularx}{\textwidth}{
    p{1.5cm}
    p{1.3cm}
    p{1.3cm}
    *{3}{>{\centering\arraybackslash}X}
}
\toprule
\textbf{Metric} & \textbf{RAMP\textsubscript{COT}} & \textbf{RAMP\textsubscript{SCoT}} & \textbf{RAMP\textsubscript{w/o I/O spec.}} & \textbf{RAMP\textsubscript{w/o t\&m limit}} & \textbf{RAMP\textsubscript{w samples}} \\
\midrule
Pass@1 & 66.5\% & 64.4\% & 56.0\% & 64.7\% & 63.8\% \\
\bottomrule
\end{tabularx}
\caption{Comparison of code generation prompting strategies and ablation studies in RAMP.}
\label{tab:prompting_overview}
\end{table}

\begin{figure}[h]
    \centering
    \begin{subfigure}{0.48\columnwidth}
        \centering
        \includegraphics[width=\linewidth]{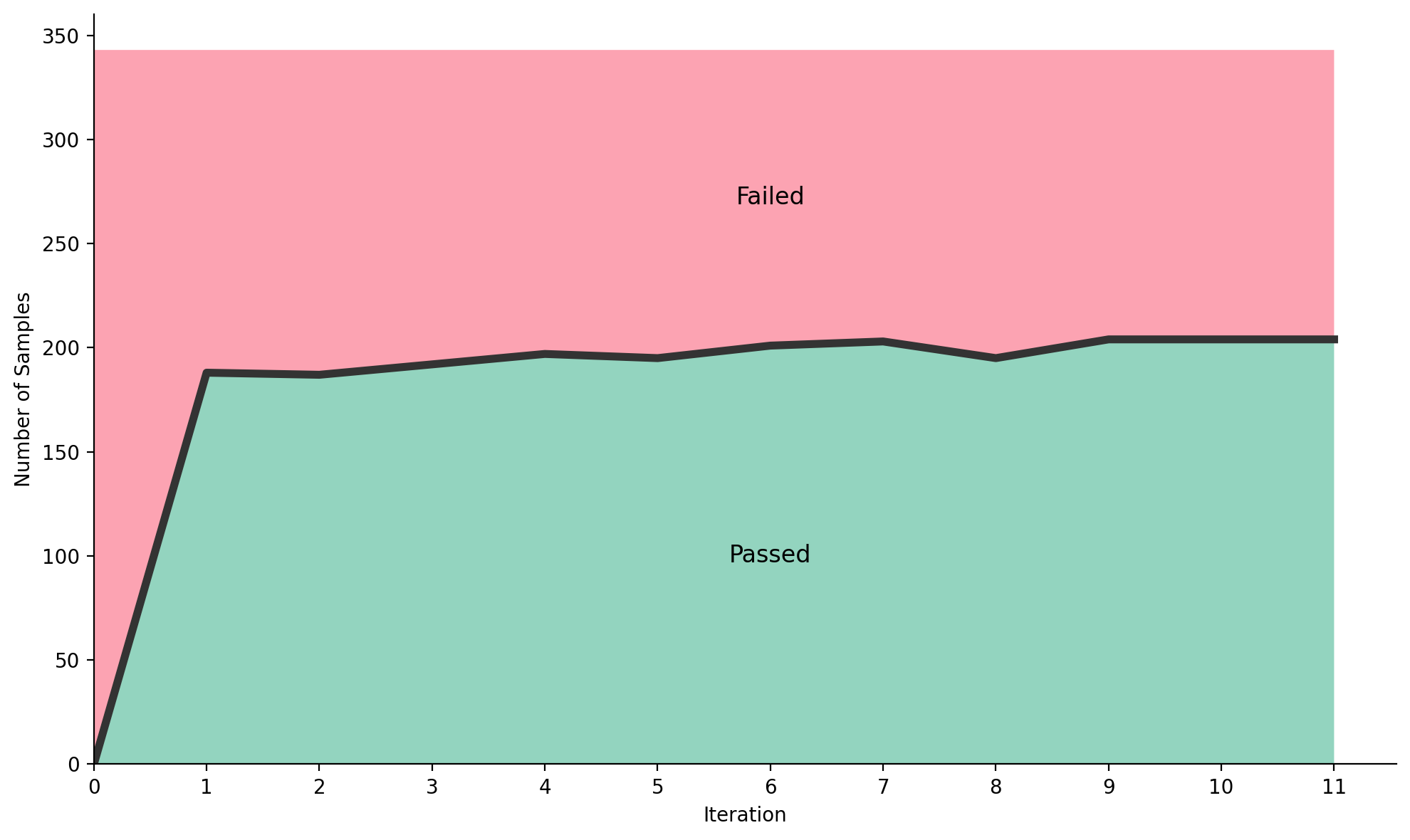}
        \label{fig:snacky}
    \end{subfigure}
    \hfill
    \begin{subfigure}{0.48\columnwidth}
        \centering
        \includegraphics[width=\linewidth]{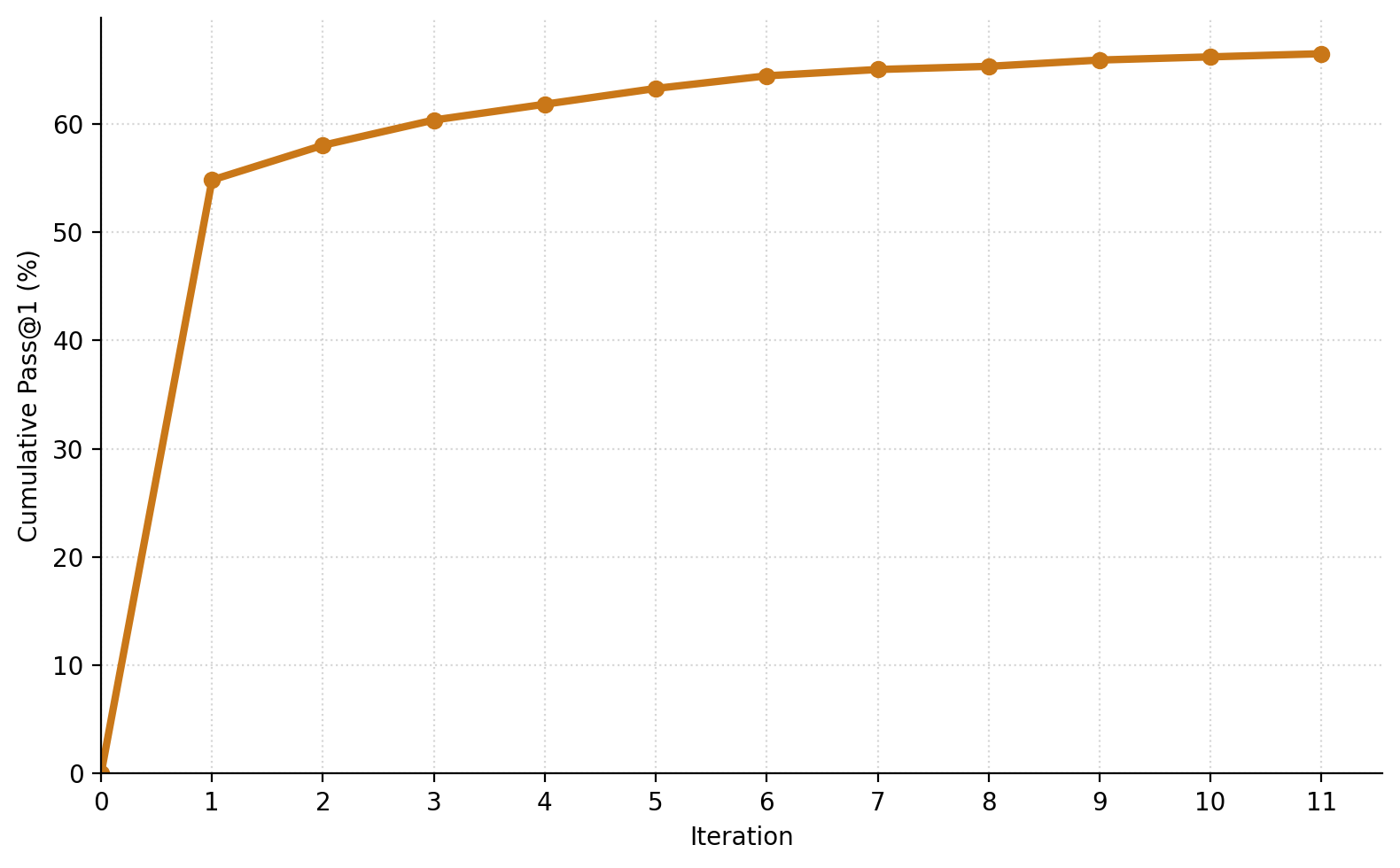}
        \label{fig:pass_at_1_cumulative}
    \end{subfigure}
        \caption{\textbf{Left:} Distribution of solved and unsolved tasks across eleven repair iterations. 
        The green region indicates the number of tasks that passed all unit tests, while the red region shows tasks that remained unsolved. 
        % The black boundary line marks the separation between the two outcomes.
        \textbf{Right:} Cumulative pass@1 across iterations. }
    \label{fig:side_by_side}
\end{figure}

Figure~\ref{fig:side_by_side} reports performance across eleven repair iterations, where at each step the generated code is evaluated against the hidden tests to track its evolution.
The left plot of Figure~\ref{fig:side_by_side} presents the evolution of solved and unsolved tasks over eleven repair iterations. The green shaded area corresponds to the 
number of tasks successfully passing all hidden unit tests, whereas the red area shows those that remain unsolved. The solid black line represents the number of solved tasks at each iteration. Notably, the line does not always increase: in some cases, tasks that were initially repaired become incorrect again in subsequent iterations. This highlights that iterative repair strategies may introduce regressions, where new modifications break previously working solutions. 
While additional iterations bring occasional improvements, the overall pattern 
demonstrates a performance plateau and instability in the repair process.
This suggests that simply increasing the number of iterations does not guarantee progress, as regressions may reverse earlier improvements.

The right plot of Figure \ref{fig:side_by_side} shows the cumulative pass@1 accuracy across eleven repair iterations. 
% Cumulative accuracy here refers to the total number of tasks that have been solved at least once up to a given iteration, summed across all prior iterations. 
As expected, the curve starts at $0$ in iteration 0.0, since no samples are solved before any repair attempt. After the first iteration, the cumulative pass@1 sharply increases to more than $50$ points, indicating that a large fraction of tasks are solved immediately with the help of the first reflection. Across iterations, cumulative accuracy improves steadily, converging to a plateau of 66.5\% by iteration 11.
% \FHF{What do you mean by cumulative accuracy? }
This trend highlights two key observations: (1) the majority of solvable tasks are captured in the first iteration, and (2) while additional iterations provide incremental gains, their impact diminishes over time. The flattening of the curve illustrates diminishing returns, suggesting that further repair attempts beyond iteration 11 are unlikely to increase overall success.

\begin{RQFinding}{Key Finding for RQ3}
Few-shot CoT prompting proved more effective than SCoT. Prompt ablations revealed that input-output specifications are critical, while time and memory limits also contribute modestly. 
% Adding sample I/O pairs, however, reduced performance by 2.7\%. 
Iterative repair analysis shows that most solvable tasks are addressed in the first iteration, with later iterations offering diminishing returns and occasional regressions, ultimately plateauing at 66.5\% pass@1 by iteration 11.
\end{RQFinding}

% \begin{RQFinding}{Key Finding for RQ3}
% Few-shot CoT prompting proved more effective than SCoT, yielding a 2.1\% higher pass@1 and becoming our default strategy. Prompt ablations revealed that input–output specifications are critical (removal reduces pass@1 by 10.5\%), while time and memory limits also contribute modestly (+1.8\%). Adding sample I/O pairs, however, unexpectedly reduced performance by 2.7\%. Iterative repair analysis shows that most solvable tasks are addressed in the first iteration, with later iterations offering diminishing returns and occasional regressions, ultimately plateauing at 66.5\% cumulative pass@1 by iteration 11.
% \end{RQFinding}

\subsection{RQ4: RAMP Performance Across Various Difficulty Levels, Subject Domains, and Execution Outcomes}

\begin{figure}[h]
    \centering
    \begin{subfigure}{0.48\columnwidth}
        \centering
        \raisebox{1cm}{\includegraphics[width=\linewidth]{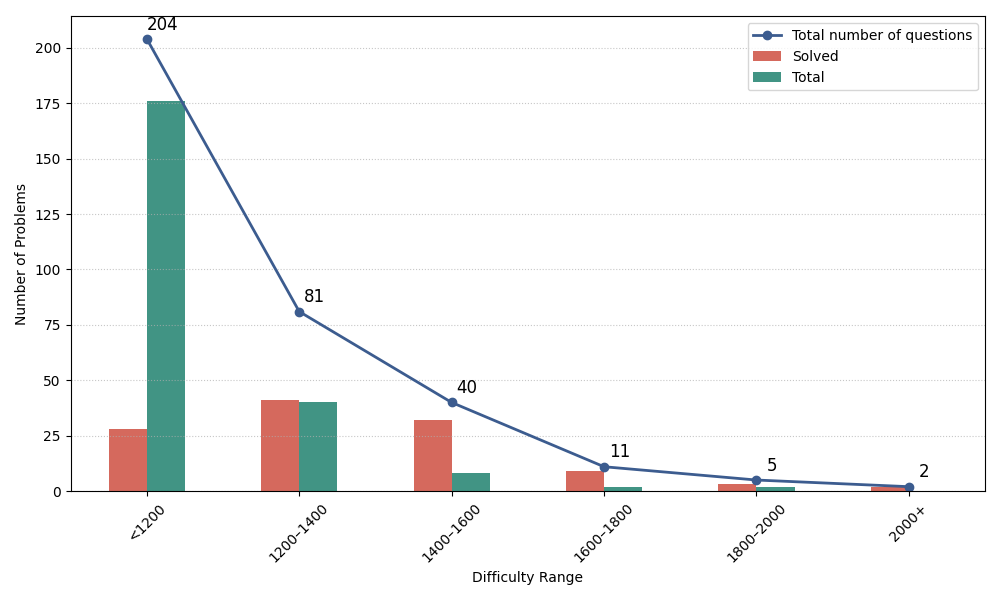}}
        \label{fig:difficulty}
    \end{subfigure}
    \hfill
    \begin{subfigure}{0.48\columnwidth}
        \centering
        \includegraphics[width=\linewidth]{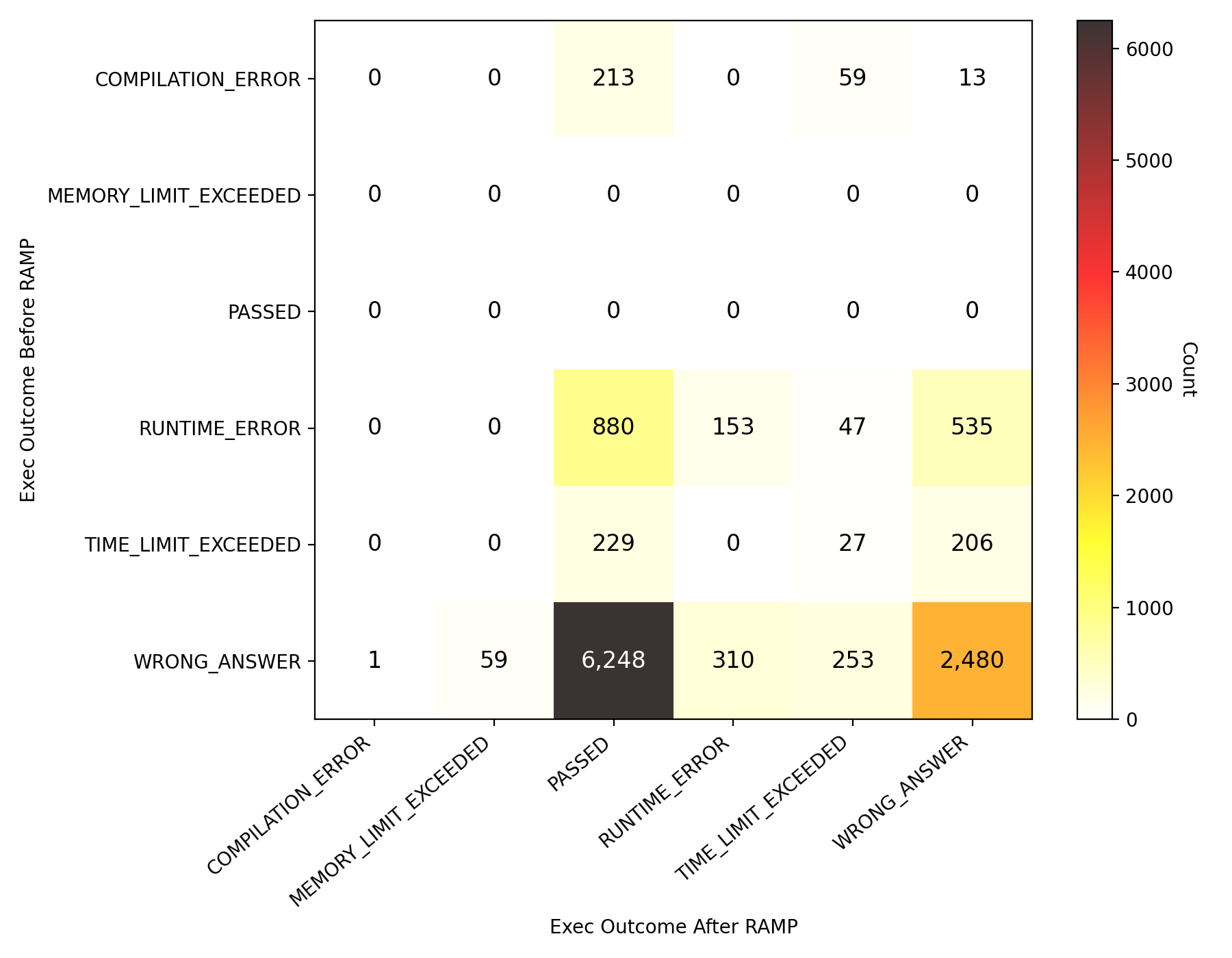}
        \label{fig:bug_outcome}
    \end{subfigure}
    \caption{\textbf{Left:} Solved and unsolved problems in RAMP across difficulty ranges. The blue line shows the number of questions in each difficulty range.
    \textbf{Right:} Bug execution outcome before and after RAMP.}
    \label{fig:difficulty_bug_outcome}
\end{figure}

The left plot of Figure~\ref{fig:difficulty_bug_outcome} presents the distribution of solved and 
unsolved problems after applying RAMP across different difficulty ranges. The difficulty ranges are represented in the metadata of the benchmark dataset, where higher numbers show a more difficult problem. Note that the number of problems in the benchmark dataset decreases as the difficulty level increases. 
The majority of solved instances fall within the lowest difficulty bracket ($<1200$), where the model achieves strong performance, solving more than five times as many problems as it fails. Performance is balanced in the 1200--1400 range, with nearly equal numbers of solved and unsolved problems, while in the 1400--1600 range, the number of unsolved problems clearly dominates. For tasks with difficulty above 1600, the success rate drops sharply, with very few problems being solved beyond this threshold.  
This distribution highlights a strong dependency between task difficulty and repair success. The model is effective at repairing simple problems, moderately effective at medium-difficulty tasks, but struggles on harder questions.

Next, we analyze how execution outcomes evolve after applying \textsc{RAMP}. 
In the \textsc{xCodeEval} benchmark, each problem is associated with a single execution outcome, recorded prior to repair. 
However, after applying \textsc{RAMP}, each hidden unit test may result in a distinct execution outcome. 
To enable a fair comparison, we assume that all hidden test cases share the same pre-repair outcome reported in the benchmark. 
The right plot of Figure~\ref{fig:difficulty_bug_outcome} presents the transition matrix of execution outcomes before and after applying \textsc{RAMP}. 
The most notable trend is that a substantial portion of programs that initially resulted in \texttt{WRONG\_ANSWER} were successfully repaired and transitioned to \texttt{PASSED} ($6{,}248$ cases). 
Nevertheless, a considerable number of these instances remained unsolved, still producing \texttt{WRONG\_ANSWER} outcomes ($2{,}480$ cases). 
Among the other categories, programs with \texttt{RUNTIME\_ERROR} were the second most successfully repaired, with $880$ cases transitioning to \texttt{PASSED}. 
Interestingly, a large fraction of these also shifted into new error categories, most prominently \texttt{WRONG\_ANSWER} ($535$ cases) and \texttt{RUNTIME\_ERROR} ($153$ cases). 
Finally, problems that originally failed with \texttt{COMPILATION\_ERROR} also showed improvements, with $213$ cases moving to \texttt{PASSED} and smaller proportions shifting to \texttt{TIME\_LIMIT\_EXCEEDED} ($59$ cases) or \texttt{WRONG\_ANSWER} ($13$ cases). 

Overall, these results highlight that \textsc{RAMP} is particularly effective at repairing \texttt{WRONG\_ANSWER} and \texttt{RUNTIME\_ERROR} cases, while also mitigating a subset of \texttt{COMPILATION\_ERROR} failures. 
At the same time, transitions into new error categories demonstrate that repair attempts may introduce different types of execution failures, suggesting opportunities for further refinement of the framework.

\begin{figure}[h]
    \centering
    \begin{subfigure}{0.79\textwidth}
        \centering
        \includegraphics[width=\linewidth]{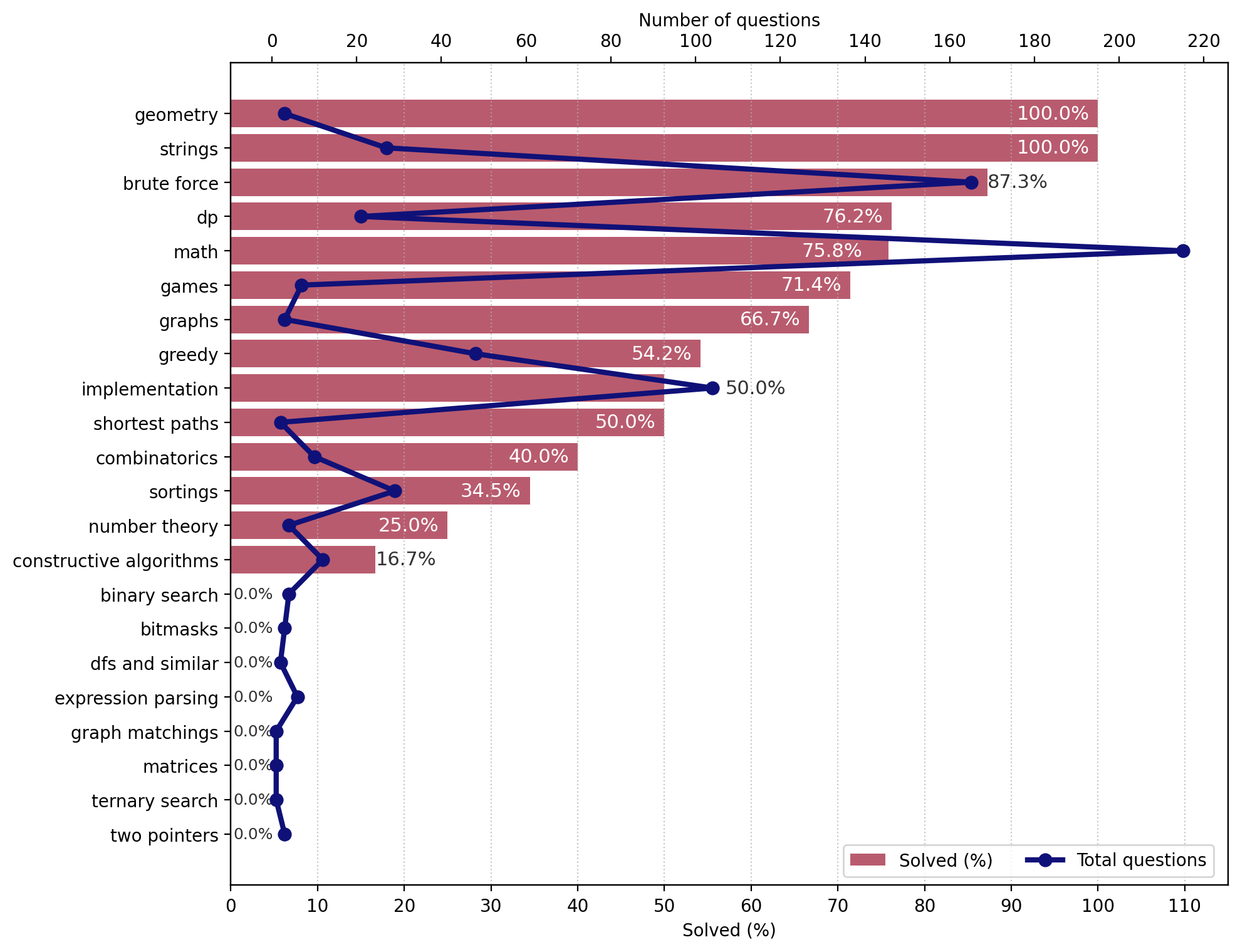}
        % \caption{Percentage of solved questions for each tag.}
        \label{fig:tags}
    \end{subfigure}
    \hfill
    \begin{subfigure}{0.20\textwidth}
        \centering
        \rotatebox{90}{\includegraphics[height=0.75\linewidth]{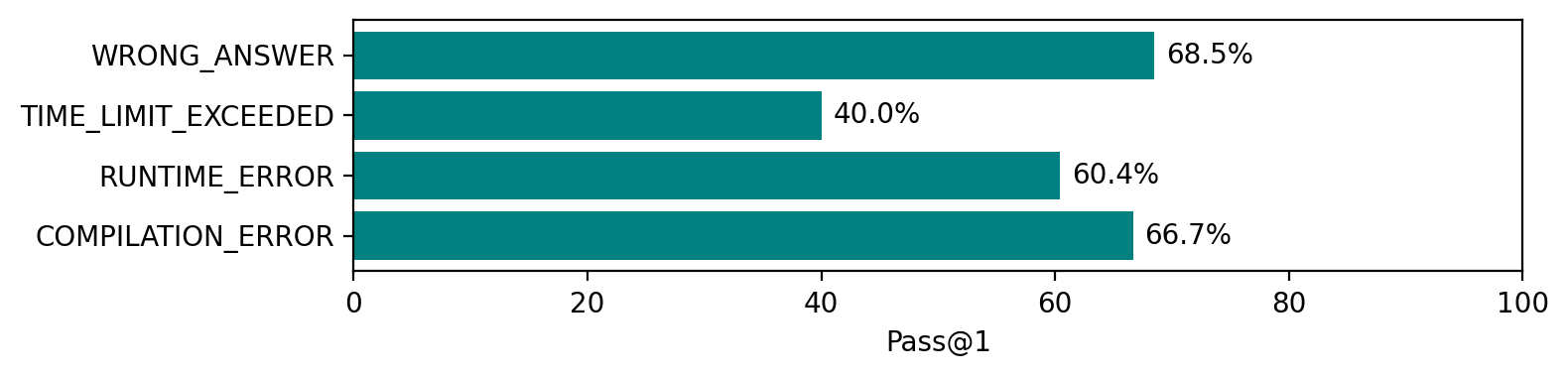}}
        % \caption{Pass@1 for each bug execution outcome.}
        \label{fig:pass1_bug_outcome}
    \end{subfigure}
    \caption{\textbf{Left}: Percentage of solved questions for each tag. The blue line shows the number of problems in each tag. \textbf{Right}: Pass@1 for each bug execution outcome.}
    \label{fig:tags_bug_outcome}
\end{figure}

The right plot of Figure~\ref{fig:tags_bug_outcome} reports the \texttt{Pass@1} score across different bug execution outcomes in the benchmark. 
As there were not any \texttt{MEMORY\_LIMIT\_EXCEEDED} in bug execution outcomes for Ruby, it is not shown in this figure.
Note that in Figure~\ref{fig:difficulty_bug_outcome} (right plot), the bug execution outcomes per test case is shown, while here, the outcome per problem in the benchmark dataset is represented. Furthermore, as we consider greedy approach for calculating Pass@1, the score also shows the percentage of the problems being solved after applying RAMP. 

We observe that \textsc{RAMP} is most effective for repairing programs that initially produced \break \texttt{WRONG\_ANSWER}, achieving a success rate of $68.5\%$. 
This suggests that although \texttt{WRONG\_ANSWER} is the most frequent failure mode before repair (cf. Figure~\ref{fig:difficulty_bug_outcome}), \textsc{RAMP} is particularly capable of correcting such cases. 
The second-highest success rate is observed for \texttt{COMPILATION\_ERROR} outcomes, with $66.7\%$ of these programs being successfully repaired. 
Programs that failed with \texttt{RUNTIME\_ERROR} also benefited substantially, with a Pass@1 of $60.4$\%. 
By contrast, programs suffering from \texttt{TIME\_LIMIT\_EXCEEDED} proved more challenging to repair, with only $40.0\%$ achieving a successful repair. 
These results highlight that while \textsc{RAMP} is highly effective at repairing common outcome categories such as \texttt{WRONG\_ANSWER}, its performance varies depending on the nature of the execution failure. 
In particular, failures due to resource constraints (\texttt{TIME\_LIMIT\_EXCEEDED}) appear less amenable to repair.
% , suggesting that future work could focus on strategies specifically designed to address performance-related bugs.

To better understand the strengths and limitations of \textsc{RAMP}, we examine which categories of problems it can solve. 
Each problem in the benchmark is annotated with one or more tags that describe its underlying topic or algorithmic technique. 
We use this information to report the percentage of solved questions for each tag, as shown in Figure~\ref{fig:tags_bug_outcome}, left plot. 
When a question has multiple tags, we count it once per tag. 
We find that \textsc{RAMP} achieves perfect success on problems labeled with \texttt{geometry} and \texttt{strings}, where $100\%$ of the questions were solved. 
Following these, the highest success rates are observed for \texttt{brute force}, \texttt{dynamic programming (dp)}, \texttt{math}, \texttt{games}, and \texttt{graphs}, with more than $60\%$ questions solved in each of these categories. 
In contrast, categories such as \texttt{greedy}, \texttt{implementation}, and \texttt{shortest paths} show moderate success rates around $45\%-55\%$, while tags such as \texttt{combinatorics}, \texttt{sortings}, and \texttt{number theory} fall below $50\%$. 
Finally, several specialized categories remain unsolved: \texttt{binary search}, \texttt{bitmasks}, \texttt{dfs and similar}, \texttt{expression parsing}, \texttt{graph matchings}, \texttt{matrices}, \texttt{ternary search}, and \texttt{two pointers}. 
% \FHF{Add a few sentences discussing the number of problems too. For eaxmple, though only 75 percent of math problems is solved, it has around 220 problems. This is still good. For the 0 percent solved problems, they are less than 10. Also discuss the reasons of ramp being able to solve some question types, not the rest. }

It is also important to consider the number of problems in each category. For instance, while the success rate for \texttt{math} is about 76\%, this corresponds to more than 200 questions, showing that \textsc{RAMP} can handle a large and diverse set of mathematically oriented tasks. By contrast, the categories with 0\% solved are extremely small, each containing fewer than 10 questions, so their impact on overall performance is limited.
These results suggest that \textsc{RAMP} is particularly effective on broad, well-represented categories such as \texttt{geometry}, \texttt{strings}, \texttt{math}, \texttt{dp}, and \texttt{brute force}, where recurring patterns are easier to capture. In contrast, its performance drops sharply on advanced paradigms and niche categories such as bitmasks or graph matchings, which demand precise reasoning and domain-specific knowledge that remain challenging for current LLM-based repair methods.

% After examining the results, since we saw that the LLM is not paying attention to self-reflection, we changed the prompt for function generation to this:
% ``Before writing the improved implementation, please answer:"
% "1. What specific changes are you going to make to the code based on the reflection?"
% "2. How exactly will this change address the issue?"
% "Then, write your full improved implementation in Ruby."
% "Ensure that your code actually reflects the reasoning above and addresses the problem."
% Pass@1 is 0.31  \texttt{check\_edge\_first\_edgeIO\_CoTIO\_CoT}.COT

\begin{RQFinding}{Key Finding for RQ4}
\textsc{RAMP}’s performance is strongly influenced by task characteristics. It excels on easier problems but struggles as difficulty increases. By execution outcome, \textsc{RAMP} is most effective at repairing \texttt{WRONG\_ANSWER} (68.5\%), followed by \texttt{COMPILATION\_ERROR} (66.7\%) and \texttt{RUNTIME\_ERROR} (60.4\%), while resource-related failures (\texttt{TIME\_LIMIT\_EXCEEDED}, 40.0\%) remain challenging. Domain-wise, \textsc{RAMP} achieves perfect repair on \texttt{geometry} and \texttt{strings}, and strong performance on \texttt{brute force}, \texttt{dp}, \texttt{math}, and \texttt{graphs}, but fails on advanced or niche categories.
% such as \texttt{binary search}, \texttt{bitmasks}, and \texttt{matrices}.
\end{RQFinding}

\section{Discussion} \label{sec:discussion}

\textbf{COT vs SCoT.} Our experiments show that SCoT lowered pass@1 compared to CoT (See section \ref{RQ3}). While the original SCoT work reported improvements \cite{li2025structured}, other studies in code generation have also observed cases where SCoT reduces accuracy \cite{zhang2024pair}. This suggests that the benefits of structured reasoning may vary significantly across tasks. In the context of program repair, where solutions often require small, context-sensitive edits, rigid structures can constrain the model’s flexibility, making free-form CoT more effective.

\textbf{Problem Difficulties.} Previous studies show that although Core Ruby Concepts are essential and widely used, they are also rated as the most difficult area, with 31.6\% of developers finding them challenging \cite{akbarpour2025unveiling}. Arrays in Ruby are the most popular group of Ruby questions on StackOverflow, illustrating that even fundamental operations can pose difficulties in practice. These difficulties align with the algorithmic problems in the \textsc{xCodeEval} benchmark, where our framework performs strongly, particularly on easy and moderate tasks. By repairing exactly the types of errors that developers struggle with most (reported in \cite{akbarpour2025unveiling}), RAMP provides practical and meaningful automated support. At the same time, the persistent difficulty of harder problems highlights an opportunity for future research, where extending and building upon RAMP could advance automated repair methods for increasingly complex cases.

\textbf{Test-based Early Stopping.} In the practical \textsc{RAMP} workflow, repair proceeds until either the iteration budget is exhausted or the candidate satisfies the generated tests. However, as Table~\ref{tab:test_confusion} shows, roughly half of the generated tests are false negatives, so they are weak stopping signals. Coupled with the plateau in Figure~\ref{fig:side_by_side}, where pass@1 shows little gain beyond a few iterations, we can adopt a hidden-test early-stopping rule: after each code generation, execute the candidate once against the hidden tests; if it passes, terminate further iterations for that sample. To indicate footprint, we report instantaneous end-of-run utilization snapshots: continuing all iterations yields 24{,}229.25~MB GPU / 4.5\% CPU / 1{,}748.66~MB RAM, whereas early stopping yields 29{,}269.25~MB GPU / 10.4\% CPU / 2{,}215.67~MB RAM. These point-in-time readings should be interpreted as lower bounds. The trade-off is a longer wall-clock time (about \(1.3\times\)) due to the added acceptance checks.

\begin{figure}[h]
    \centering
    \begin{subfigure}{0.49\columnwidth}
        \centering
        \includegraphics[width=\linewidth]{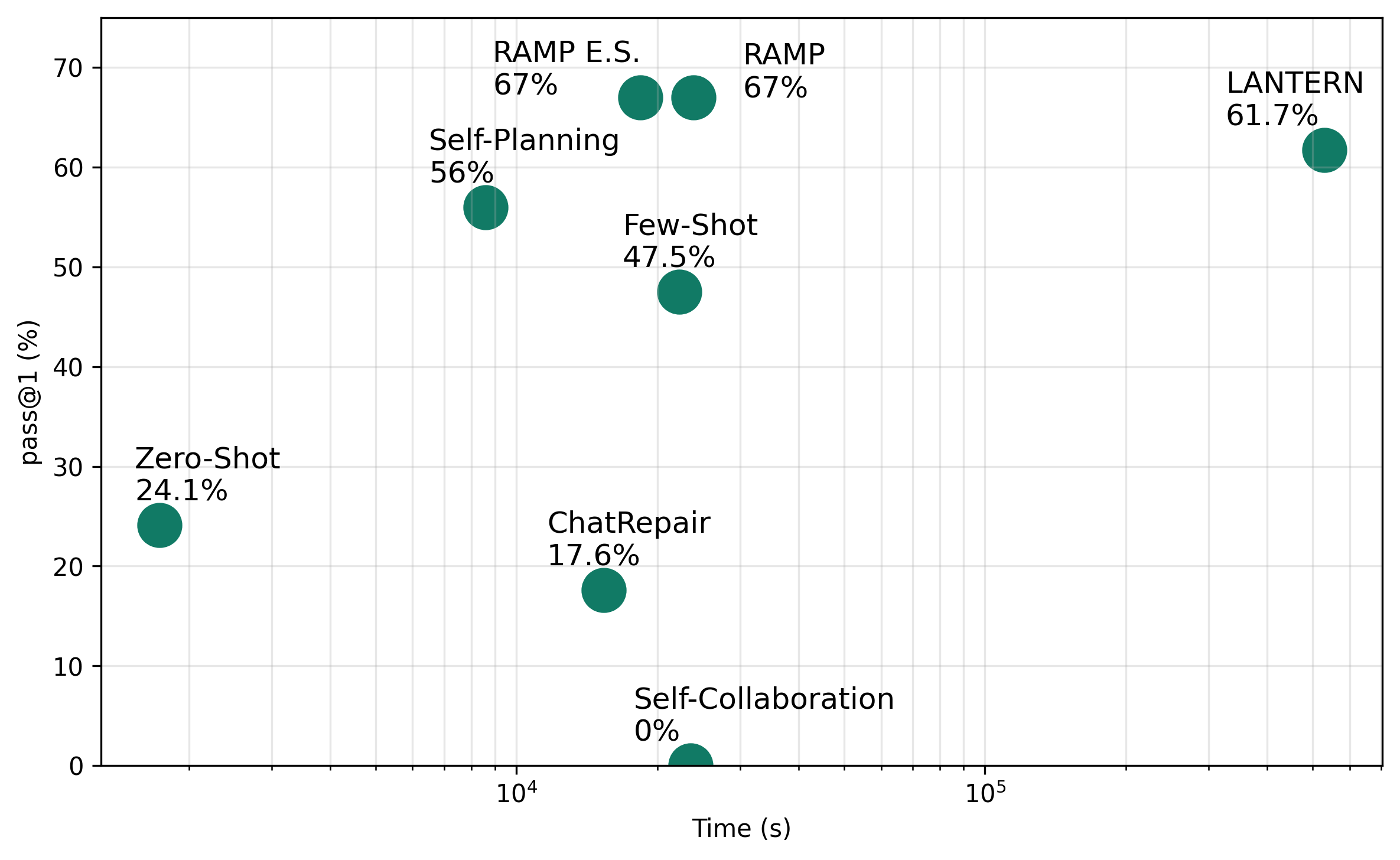}
        \label{fig:pass_vs_time}
    \end{subfigure}
    \hfill
    \begin{subfigure}{0.48\columnwidth}
        \centering
        \includegraphics[width=\linewidth]{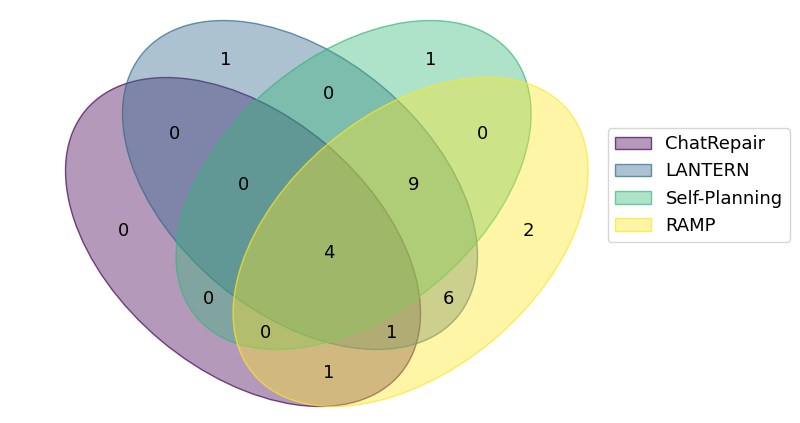}
        \label{fig:venn}
    \end{subfigure}
        \caption{The \textbf{left} plot shows the pass@1 and time consumed of RAMP compared to baselines. \textsc{RAMP E.S.} is RAMP with early stopping rule. 
        The \textbf{right} plot shows bug fix Venn diagram on \textsc{xCodeEval}.}
    \label{fig:pass_vs_time_venn}
\end{figure}

\textbf{Performance vs. Time.}
Figure~\ref{fig:pass_vs_time_venn} (left) compares end-to-end runtime on the 10\% Ruby validation subset of the validation set against pass@1. 
\textsc{RAMP} achieves the highest pass@1 with a moderate runtime (67 in $\approx 2.4\times10^{4}$\,s, $\sim$6.6\,h), yielding the strongest overall balance of accuracy and time.
\textsc{RAMP E.S.}, a variant with an early stopping rule, consumes more GPU resources but finishes faster ($\approx 1.8\times10^{4}$\,s, $\sim$5.1\,h).
Self-Planning is the most efficient baseline (56 in $\approx 8.6\times10^{3}$,s, $\sim$2.4,h), while Few-Shot offers mid-range performance (47.5 in $\approx 2.3\times10^{4}$,s, $\sim$6.4,h).
Zero-Shot is very fast ($\approx 1.7\times10^{3}$,s) but low pass@1 (24.1\%), whereas \textsc{ChatRepair} is slower yet less accurate (17.6\%).
Self-Collaboration adds time without benefit (0.0\%).
All points except \textsc{LANTERN} report time measured on the Ruby subset of the validation set. \textsc{LANTERN} requires building a cross-language database before repair, so it cannot be run on Ruby alone. The reported \textsc{LANTERN} time, therefore, includes (i) constructing the database over the entire 10\% multilingual validation split and (ii) repairing all languages, which explains its substantially larger runtime (61.7 in $\approx 5.3\times10^{5}$\,s, $\sim$147\,h). 
Overall, for Ruby-only repair under comparable compute, \textsc{RAMP} offers the best accuracy-time trade-off. Self-Planning is the most efficient alternative, whereas Zero-Shot, \textsc{ChatRepair}, and Self-Collaboration provide little value for their time, and cross-language systems like \textsc{LANTERN} are inefficient for this scope due to the heavy database-construction and multi-language repair overhead.

\textbf{Overlapped Solved Problems.}
Figure~\ref{fig:pass_vs_time_venn} (right) shows that solved problems are dominated by \textsc{RAMP}, which achieves the broadest overall coverage, including two unique solves, while sharing most solutions with both Self-Planning and \textsc{LANTERN}. Self-Planning and \textsc{LANTERN} each contribute one unique solve, though most of \textsc{LANTERN}’s successes overlap with \textsc{RAMP}. The intersection of all four methods yields four shared solves, and the remaining overlaps are minimal.
% (e.g., 1 for {\textsc{ChatRepair}, \textsc{Self-Planning}, \textsc{RAMP}} and 1 for {\textsc{LANTERN}, \textsc{Self-Planning}, \textsc{RAMP}}). \textsc{ChatRepair} provides no unique solves.
These patterns, together with the time-accuracy results, indicate that \textsc{RAMP} is the strongest single method, covering the broadest set of problems while also capturing much of the overlap with other approaches. Self-Planning remains a complementary, lightweight option but contributes little unique coverage, while \textsc{LANTERN} overlaps heavily with \textsc{RAMP} 
% despite adding only 1 unique solves and 
while requiring costly preprocessing.

\textbf{Practicality of RAMP.}
RAMP framework relies on feedback from the \textit{Test Designer Agent}. Removing it and utilizing the outcomes of the hidden unit tests in the benchmark raises performance on the RQ1 dataset to \textbf{91.17\% pass@1}. However, designing new tests is more realistic and closer to real-world applications, where we do not always have access to test cases for buggy code. 
Another advantage of RAMP is its simple adaptation to other programming languages. This only requires swapping the executor for one supporting the target language and updating few-shot examples. To prove this, we have tested RAMP on C++ portion of the dataset used in RQ1 (10\% of \textsc{xCodeEval} validation), which has 65 samples. RAMP outperforms other baselines on C++ by a big margin, achieving pass@1 of \textbf{32.3\%}. This is compared to pass@1 of 23.0\%, 7.0\%, 20.4\%, and 7.0\% for LANTERN, \textsc{ChatRepair}, Self-Planning, and Self-Collaboration, respectively.

\section{Threats to Validity}

\paragraph{Internal Validity}
Our results may be influenced by several internal factors. First, RAMP relies on LLM-generated test cases for intermediate feedback; these test cases may not always be correct, which can affect the quality of guidance during repair. To mitigate this, we use generated tests only as auxiliary signals and never for final evaluation. Second, the behavior of individual agents, particularly the \textit{Programmer} and \textit{Feedback Integrator}, is sensitive to prompt design; small variations in wording or structure can yield different outcomes. To reduce prompt-induced variance, we first try different prompts and use the same prompt template across all runs and baselines. Third, our choices of different parameters (e.g., temperature, top-p) can introduce internal bias. We mitigate this by adopting best-reported settings from prior work or model providers.
% Finally, model variance plays a role: decoding strategies (e.g., temperature, random seeds) introduce nondeterminism, which may affect reproducibility across runs.

\paragraph{External Validity}
The generalizability of our findings is subject to certain limitations. We conducted experiments on the \textsc{XCodeEval} benchmark, which covers competitive-programming-style tasks. These problems may not fully reflect the complexity or diversity of real-world Ruby projects.
Moreover, although \textsc{RAMP} was primarily designed and evaluated for Ruby, we also included a limited evaluation on C++. The framework is theoretically applicable to other programming languages, but we did not systematically investigate this aspect in the present study.

\paragraph{Construct Validity}
% We evaluate repair effectiveness using \textit{Pass@1} on hidden unit tests. While this is a widely adopted metric, passing all provided tests does not guarantee semantic correctness or maintainability of the produced patches. Some repairs may overfit to the test suite.
We have used benchmarks and evaluation metric that are widely used \cite{jiao2024preference, islam2024mapcoder, luo2025unlocking} to avoid any threats.
Our analysis by difficulty level assumes that benchmark-provided scores are accurate proxies for problem complexity, which may not always hold. To reduce this risk, we use these scores only to group problems and report overall results so the analysis can be repeated with different difficulty definitions.
A second threat is our choice of evaluation metric. We primarily report \textit{pass@1} based on hidden unit tests, which may miss qualities such as efficiency and readability. We address this by providing our evaluation script and per-task identifiers so others can re-score with alternative metrics or test suites.
Additionally, due to resource constraints, RQ1 evaluations were performed on a 10\% subset of the validation set. To avoid bias, our sampling preserved the original language and difficulty distributions.
% Furthermore, while our ablation studies isolate the contribution of different agents, interactions among components mean that their effects cannot be completely disentangled.

\paragraph{Conclusion Validity}
Our conclusions are based on measured pass@1 accuracy on the \textsc{XCodeEval} benchmark. While our experiments show consistent improvements with RAMP, some analyses involve small groups of problems (e.g., specific tags or error categories), which means the percentages reported there may not fully reflect general trends. Therefore, our conclusions about overall effectiveness are reliable, but more fine-grained claims should be interpreted with caution.

\section{Conclusion and Future Work}

We introduced \textsc{RAMP}, a lightweight multi-agent framework for Automated Program Repair in Ruby. By structuring repair as a feedback-driven process, \textsc{RAMP} leverages test generation and self-reflection to iteratively refine candidate solutions. Evaluation on the \textsc{XCodeEval} benchmark demonstrated that \textsc{RAMP} achieves state-of-the-art performance on Ruby.
% , with a Pass@1 of 67\%, outperforming other baselines.
Our analysis further showed that \textsc{RAMP} converges within five iterations, maintains efficiency compared to resource-heavy methods, and is particularly effective on broad and well-represented problem categories.
% , while still struggling with specialized algorithmic domains. 
Together, these findings highlight the promise of multi-agent reasoning for practical, efficient program repair beyond traditionally studied languages.
Future directions for this research includes enhancing domain-specific reasoning and improving the reliability of the generated tests that could further strengthen \textsc{RAMP}’s iterative repair loop. 

\section{Data Availability}
Full replication package and experimental data are available at \break \url{https://figshare.com/s/829875edc8c876c50de5}.

% \section*{References}

\bibliographystyle{ACM-Reference-Format}
% \bibliography{references}

\end{document}